\documentclass[twocolumn,aps,superscriptaddress]{revtex4-2}
\usepackage[dvipsnames]{xcolor}
\usepackage{amsmath}
\usepackage{amsfonts}
\usepackage{amssymb}
\usepackage{hyperref}
\usepackage{graphicx}
\usepackage{multirow}
\usepackage{comment}

\begin{document}
\title{Complete insensitivity to \textit{ab initio} data -- A new perspective on modeling collision-induced absorption of noble gas atoms }

\author{Nikhila A. Chandran}
\affiliation{Institute for Molecules and Materials, Radboud University, Nijmegen, The Netherlands}
\author{Tijs Karman}
\email{tkarman@science.ru.nl}
\affiliation{Institute for Molecules and Materials, Radboud University, Nijmegen, The Netherlands}

\begin{abstract}
 In this study, we systematically investigate how the accuracy of collision-induced absorption (CIA) spectra of noble gas atom pairs depends on the quality of the \textit{ab initio} data used. We evaluate quantitatively the impact of different quantum chemical methods and basis sets on spectral features, finding that even the lowest-level calculations are accurate to approximately 10~\% at room temperature, and better at higher temperatures.
 This study also reveals a previously unreported double-peak structure for the He-Ne complex, which cannot be described by simple but commonly used single-exponential models for the short-range dipole. Our analysis shows that the range of internuclear distances relevant for CIA spectra varies with temperature, with short-range interactions becoming increasingly important at high temperatures. The long-range van der Waals induced dipoles never contribute. These findings provide new insights into the temperature dependent behavior of CIA spectra and emphasize the importance of accurate modeling of short-range interactions for reliable astronomical modeling.
\end{abstract}

\maketitle
\section{Introduction}
When gas molecules collide, their normal charge distribution distorts, resulting in an interaction-induced dipole. This induced dipole allows the complex to absorb electromagnetic radiation in the microwave and infrared frequencies, producing collision-induced absorption (CIA) spectra\cite{cia_2, K_W, cia3}. CIA is significant for understanding the opacities of planets, exoplanets, and cold stars with dense atmospheres. Accurate knowledge of these opacities is necessary for astrophysical and planetary research, especially at high temperatures \cite{CIA_astronomy,vigasin1998,vigasin2003}.
The HITRAN spectroscopic database provides the parameters of CIA for astronomical modeling \cite{cia_in_hitran}.
This database shows the need for the CIA calculations of noble gas heteroatoms at various temperatures \cite{karman}. In this study, we calculate the CIA spectra for noble gas heteroatoms such as Ne--He, Ar--He, and Ar--Ne.

The intensity and shape of the spectra depend on the dipole and the potential energy surfaces of the colliding atoms. These parameters are, in turn, determined approximately to an accuracy that is limited by both the level of \emph{ab initio} methods used and the finite basis sets employed. It has often been claimed \cite{elkader,el2014spectral,el2012rototranslational,BORYSOW1984235,meyer2015,fakhardji2019collision,borysow1998computer} that CIA spectra are sensitive to both the potential and dipole surface.
However, a systematic investigation into how uncertainties in these surfaces propagate into spectral predictions has to our knowledge never been performed. Therefore, quantifying the relative impact of each parameters on the computed spectra is crucial for assessing the reliability of theoretical models. 

Traditionally, considerable emphasis has been placed on the long-range characteristics of the potential energy surface (PES) and Induced dipole surface (IDS) in spectral calculations \cite{bohr1,bohr2,fowler,heijmen1996symmetry,meyer1994long,bishop1993calculation,fowler1990d9,galatrygharbi,whisbye1,whisbye2,matchanesbet,lr2}, under the untested assumption that these features significantly influence the spectra. However, the extent to which long-range and short-range contributions influence the spectra has not been thoroughly investigated.

The short-range interactions are often approximated as deviations from the long-range form that vary exponentially with distance \cite{matchanesbet,whisbye1,whisbye2,Birnbaum,meyerfromm,McQuarrie,CIA2015,elkader,elkadersingleparameter}.
Despite the widespread use of such approximations,
the influence of this choice on the resulting spectra remains unassessed. 

This study aims to address these gaps by systematically evaluating the sensitivity of spectral predictions to variations in both the PES and IDS surfaces.
This is illustrated by calculating the CIA spectra for noble gas heteroatoms Ne--He, Ar--He, and Ar--Ne.
In particular, we quantify the role of short-range and long-range contributions,
finding that the long-range van der Waals interaction never contributes to the spectrum.
For the specific case of Ne--He, we find an unexpected structure in the spectrum,
which cannot be described if the short-range dipole moment is modeled as a monotonically decaying single-exponential.
Furthermore, we quantify the relative impact on the predicted spectrum of uncertainty in the potential and dipole surface due to \emph{ab initio} methods and basis sets.
We find that the predicted spectra are insensitive to long-range \emph{ab initio} data and the uncertainty in the spectra is approximately 10~\% with the lowest-level of electronic structure data for spectral calculations, giving a new perspective on the level of theory required for quantitatively predictive modeling of CIA spectra for reliable astronomical modeling.

\section{Electronic structure calculations}

We calculated the PESs and IDSs for Ne--He, Ar--He, and Ar--Ne in \mbox{MOLPRO}~2015 \cite{molpro15} using the RHF, CCSD, and CCSD(T) methods using augmented basis sets (aug-cc-pV$X$Z, $X = 3$,4,5,6). 
These calculations included midbond functions, with exponents defined in \cite{midbond_exp}, between the centers of the two atoms. The interaction energies are calculated using Boys and Bernardi's counterpoise correction \cite{bb}.
We extrapolated to the complete basis set (CBS) limit by fitting the  correlation energy for the two largest basis sets as $E_c(\zeta) = E_c(\infty) + c\zeta^{-3}$ \cite{cbs},
and used the Hartree-Fock energy computed from the largest basis set.
Interaction-induced dipoles are calculated using the finite field approach with a field of $\pm$0.0002 a.u. 
All the interaction energies are calculated in an equally spaced grid, $R$ (the internuclear separation between the two atoms), from 2$~a_0$ to 20$~a_0$ with a separation of 0.2$~a_0$. 
These \textit{ab initio} points are then interpolated and extrapolated using the reproducing kernel Hilbert space (RKHS) method \cite{rkhs} in a denser grid relevant to the CIA calculations.

Our recommended PESs and IDSs are computed at the CCSD(T)/CBS level.
These are shown in Fig.~\ref{pes_ids} and the data can be found in the Supplemental Material.
Calculations in the smaller basis sets and using lower levels of theory will be used below to estimate the sensitivity of the CIA spectra to the PESs and IDSs.

Figure~\ref{pes_ids}(a) shows the PESs.
For all three rare-gas pairs, the potential energy wells are shallow and increase with atomic mass.
For Ne--He, the potential well is shallow, approximately 66.3 $\mu$hartree at 5.6 $~a_0$.
For Ar--He, the well depth is 94.7 $\mu$hartree at 6.6 $~a_0$.
For Ar--Ne, the well depth increases to 206.7 $\mu$hartree, also at 6.6 $~a_0$. 
These values agree well with Refs.~\cite{cybulski, lopez}. We compute bound states supported by these potential curves by sinc-function discrete variable representation (DVR) \cite{sinc_dvr}. 
We find that both Ne--He and Ar--He have one bound state, while Ar--Ne has four bound states, consistent with the findings in \cite{cybulski}.
A previous study by Cacheiro et al., shows one more bound state for all three cases, \cite{lopez}. These near-threshold bound states are not well described in our DVR calculations, but are not expected to contribute significantly to the absorption because they are long-ranged.

Figure~\ref{pes_ids}~(b) shows the IDSs.
At large separation between the collisional partners, the dipole moment vanishes.
As the atoms approach each other, their van der Waals interaction induces a dipole moment, which increases with the polarizability of the interacting atoms.
Among the three systems studied, Ar--Ne exhibits the largest induced dipole, as the larger size of the atoms makes them easily polarizable.
The sign of this dipole moment is consistently negative, with the lighter (heavier) atom placed in the positive (negative) $z$ direction. The sign of the induced dipole moment agrees with the qualitative expectation\cite{CIA_astronomy} that as the atoms approach one another, the larger atom tends to shift its negative charge into the region between the two atoms, corresponding to attractive long-range van der Waals interactions.
 As the separation decreases further, exchange repulsion becomes dominant and reduces the charge density between the atoms, causing the dipole moment to reverse direction.
We note that this reversal of sign of the dipole function occurs at 8~$a_0$ for all combinations of rare gas atoms considered, despite their differences in polarizability and effective radius.

For both the PES and IDS, first and second-order electrostatic perturbation theory give the contributions of physically meaningful long-range interactions with characteristic power-law $R$ dependence \cite{tt,lr1,lr2,fowler}.
For rare-gas pairs, the leading contribution to the potential (dipole) is of the van der Waals type and scales with $R^{-6}$ ($R^{-7}$). 

For the systems studied in this work, the long-range regime --where the van der Waals contribution to the dipole moment dominates-- corresponds to internuclear separations greater than 8$~a_0$. 
At shorter distances, the deviations from the long-range form cannot be described by a single exponential decay, which is often assumed in calculations of CIA line shapes \cite{matchanesbet,whisbye1,whisbye2,Birnbaum,meyerfromm,McQuarrie,CIA2015,elkader}.
For Ne--He, this is particularly apparent as the dipole exhibits a second sign change around $R=4.2~a_0$.
This behavior highlights the inadequacy of single-exponential models for capturing the complexity of short-range interactions in such systems. 

\begin{figure*} 
         \centering
         \includegraphics[width=0.475\textwidth]{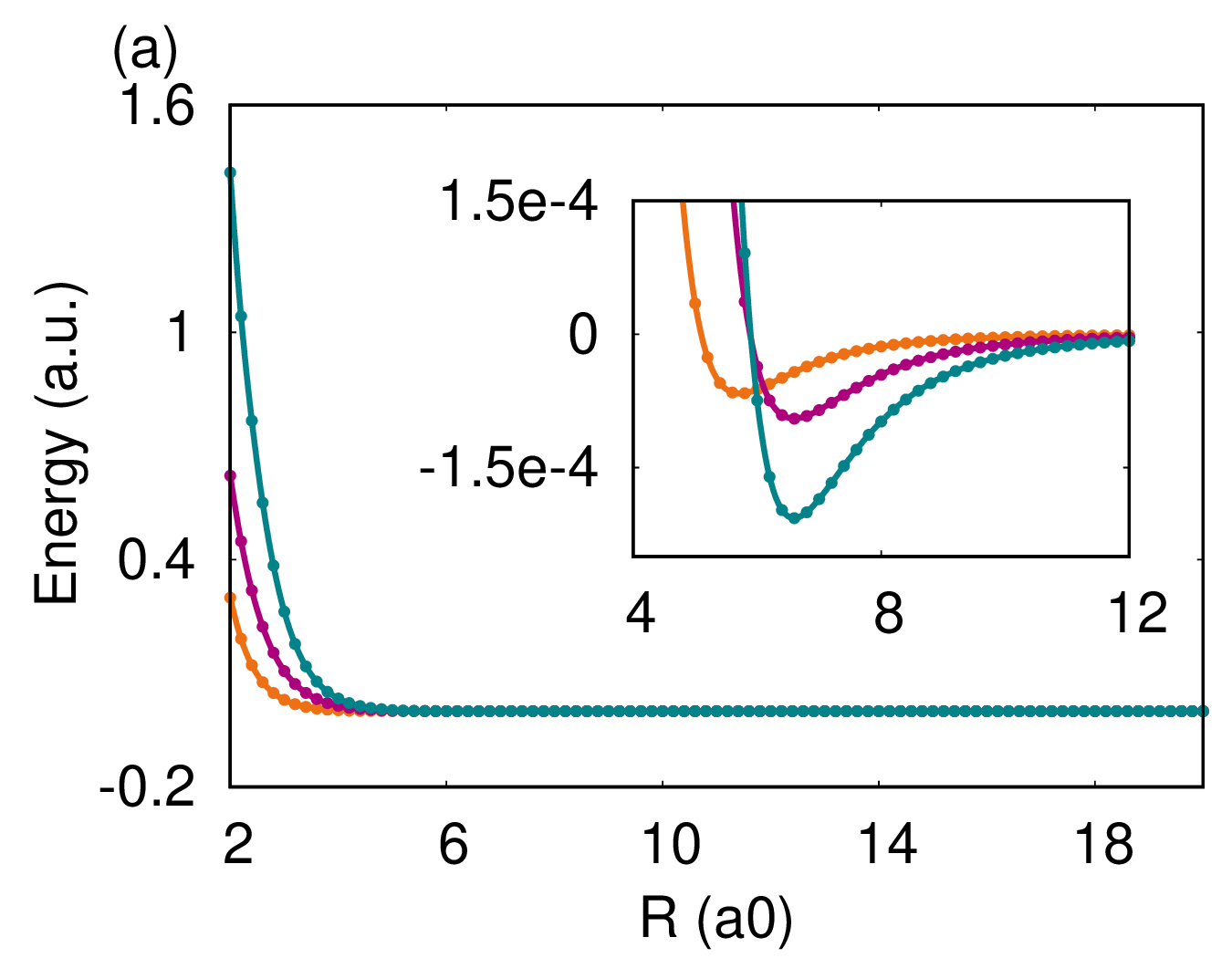}
         \includegraphics[width=0.475\textwidth]{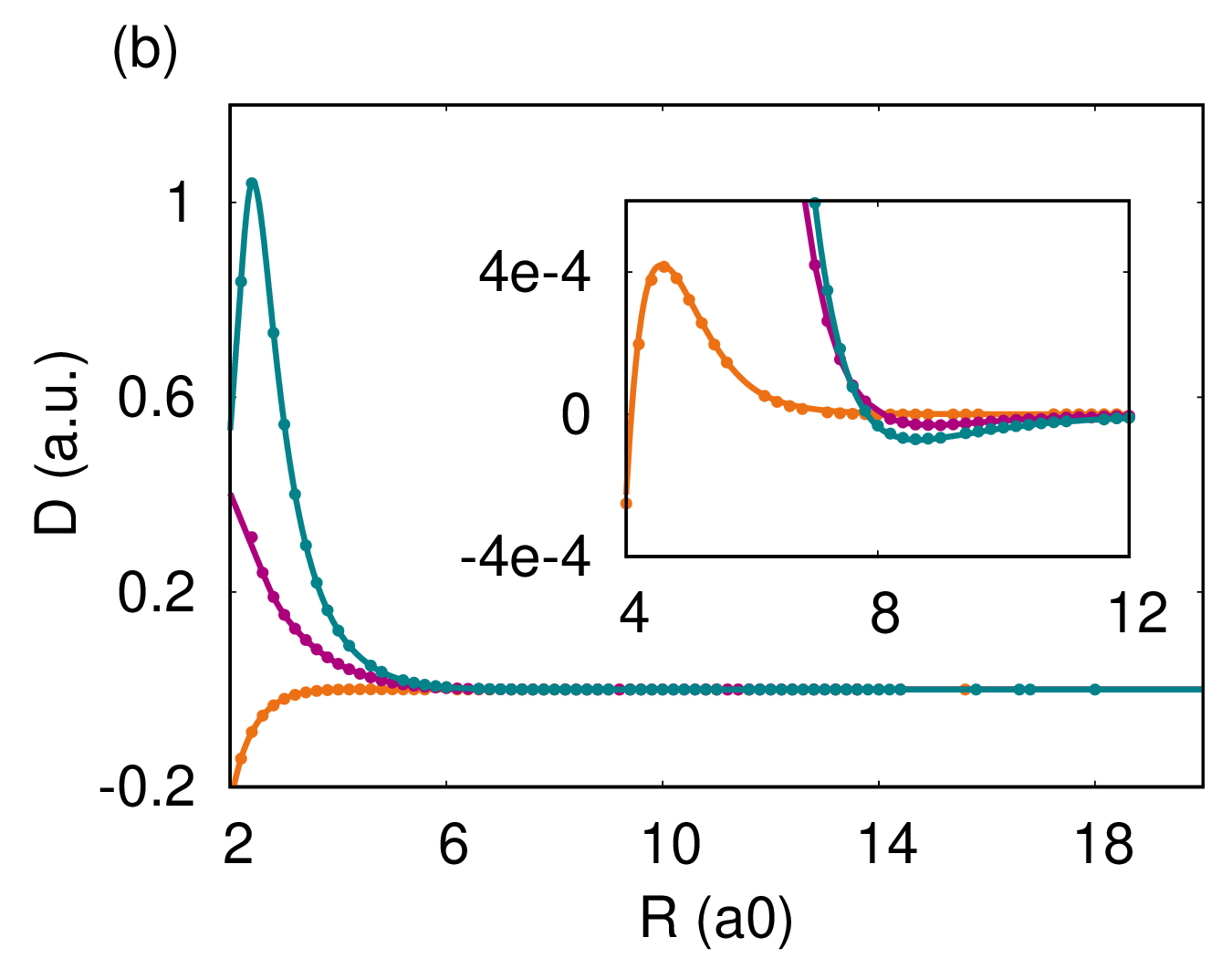}
         \includegraphics[width=0.475\textwidth]{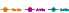}
         \caption{(a) Potential Energy Surface (PES) and (b) Induced Dipole Surface (IDS) of collisional pairs} 
         \label{pes_ids}
\end{figure*}

\section{CIA spectrum calculations}

CIA spectra of dissimilar noble gas atoms are calculated by coupling the collisional complex to the light field by first-order time-dependent perturbation theory. Spectral profiles of the absorption coefficient at various frequencies and temperatures $\alpha (\omega, T)$, normalized by the numerical densities $\rho_1$, $\rho_2$ of the dissimilar atoms, are given by \cite{Frommhold_1994}
\begin{equation} \label{CIA_eqn}
 \frac{\alpha (\omega, T)}{\rho_1 \rho_2 } = \frac{4\pi^{2}}{3 \hbar c} \omega [1-\exp(-\frac{\hbar \omega}{k_{B}T })] V g(\omega, T)
\end{equation}
with spectral density
\begin{align}
Vg(\omega,T)&  =\hbar \lambda_o^3 \sum_{L,L'} 
\begin{pmatrix}
L & \lambda & L'\\
0& 0 & 0
\end{pmatrix} ^2 \nonumber\\ 
&\quad \times \int_{0}^{\infty} dE_{col} ~\exp(-\frac{E_{col}}{k_BT}) \nonumber\\ 
&\quad \times |\langle E_{col},L| D(R)|E_{col} + \hbar \omega,L' \rangle|^2. 
  %Vg(\omega,T ) = \sumint_i \sumint_f P^{(i)} |\langle i|\hat \mu|f \rangle|^{2} \delta (\omega_f -\omega_i -\omega) .
\end{align}
The constant $c$ is the speed of light, $\hbar$ is the reduced Planck constant,
and the symbol in parentheses is a Wigner 3-$jm$ symbol.
The bra $\langle  E_{col},L|$ and ket $|E_{col} + \hbar \omega,L' \rangle$ are the energy-normalized initial and final eigenstates of the Hamiltonian without the radiation field. These wavefunctions are calculated by numerically propagating the scattering wavefunction for the Hamiltonian
\begin{align}
\hat{H} = -\frac{\hbar^2}{2\mu}\frac{d^2}{dR^2} + \frac{\hbar^2 L(L+1)}{2\mu R^2} + \mathcal{V}(R),
\end{align}
using the renormalized Numerov method \cite{tijsh2,numerov}. 

The wavefunctions are propagated in a grid of 2 to 20$~a_0$ with a grid spacing of $\Delta = \lambda_\mathrm{min}/10$, where $\lambda_\mathrm{min}$ is the shortest de Broglie wavelength for the system, at the highest temperature and wavenumber. The maximum angular momentum included in the calculations is determined by $L_\mathrm{max} = b\sqrt{2\mu E_\mathrm{max}}$, where $b=20 ~a_0$ is the maximum impact parameter, $\mu$ is the reduced mass and $E_\mathrm{max}$ is the maximum collisional energy. 
The collisional energies are taken from a logarithmically spaced grid of 31 points between $k_BT/10$ and $10k_BT$,
and all computational parameters—including grid size, spacing, $L_\mathrm{max}$ and E$_\mathrm{max}$—are systematically checked to ensure convergence of the computed spectra. 
 
 \subsection{Classical phase-space integral}
 
Quantum calculations are validated using classical statistical mechanical theory, which is computationally inexpensive \cite{csm,Frommhold_1994}. The integrated intensities at a particular temperature can be calculated using   
\begin{align}
    \int_{-\infty}^{\infty} d\omega Vg(\omega,T) &= \int_{0}^{\infty} dR~4\pi R^2 \exp(-\mathcal{V}/k_B T) D^2(R) \nonumber \\
    &= 4\pi \int_{0}^{\infty} F(R) dR,
    \label{csm}
\end{align}
where $\mathcal{V}$ is the interaction potential, $D$ is the dipole. 
Inspecting the integrand, $F(R)$, also provides insight into which ranges of $R$ contribute significantly to the spectral calculations at various temperatures.

The integrated intensity is dominated by short distances where the dipole, $D^2(R)$, is large,
but not so short that the potential is highly repulsive.
As temperature increases, the optimum of this trade off moves to shorter distances, further into the repulsive potential wall.
Deeper potential wells lead to an increase of the intensity,
explaining that the intensity grows when converging the potential with basis or method.
This sensitivity disappears when the temperature grows large compared to the potential depth.

\subsection{\label{sec:analytic}Analytical line shape model}

Having discussed the intensity, we attempt to develop a simple understanding of the line \emph{shape} here.
A first simplification of the full lineshape calculation is to approximate the thermal average over many translational states by considering only $L=0$,
and setting the collision energy equal to $k_BT$.
This gives the following approximation to the spectral density
\begin{align}
Vg(\omega,T)&  \propto |\langle k_BT,0| D(R)|k_BT + \hbar \omega,0 \rangle|^2. 
\label{softsphere}
\end{align}

A further cruder approximation is to replace the interaction potential by a hard sphere of radius $a$, which we estimate from the classical turning point for the physical potential at collision energy $k_BT$.
In this approximation, the energy-normalized radial wave function is given by
\begin{align}
  \langle R|lE_{col}\rangle = \begin{cases}
\frac{1}{\hbar R }\sqrt{\frac{\mu}{2\pi k}} \sin[k(R-a)]~ \mathrm{for}~ R \geq a\\
0     ~~~~~~~~~~~~~~~~~~~~~~~~~~~~\mathrm{for}~ R\leq a,
\end{cases}  
\end{align}
where $k$ is the wave number which satisfies $\hbar k = \sqrt{2\mu E_{col}}$.
This relates the hard-sphere approximation to the spectral density
\begin{align}
    Vg(\omega) & \propto \frac{1}{kk'} \Bigg| \int_0^\infty \sin(kR) ~\sin(k'R) D(R+a) ~dR \Bigg|^2,
   \label{hardsphere}
\end{align}
where the integral is related to the Fourier transform of the shifted dipole function, evaluated at the sum and difference of the wavenumbers $k\pm k'$.
This qualitatively means that, for short-ranged dipole surfaces, we will obtain broad absorption features, whereas, for long-ranged dipole surfaces, the absorption feature will be narrower.

Compared to the hard-sphere approximation, Eq.~\eqref{hardsphere},
we may describe our first approximation, Eq.~\eqref{softsphere}, as a soft-sphere approximation that accounts for the interaction potential.
In this case, we compute the radial wavefunctions numerically using the Numerov algorithm.
The deviations of these functions from sine waves modify the Fourier transform slightly to account for faster motion over attractive potential wells and slower motion close to the classical turning point.
Features reproduced by our simplest hard-sphere approximation can be attributed fully to the $R$ dependence of the dipole function, and can be understood in simple terms.
For example, the Fourier transform of a single exponential is a Lorentzian, which leads to a smooth absorption spectrum without local minima.
Conversely, structure at a particular frequency in the absorption spectrum --in this simplified model-- necessarily results from a modulation in the dipole function at a particular wavenumber.
Differences between the two approximations developed here, by contrast, can be related to the interaction potential.

\section{Convergence of the CIA spectra}

\begin{figure*} 
         \centering
         \includegraphics[width=0.475\textwidth]{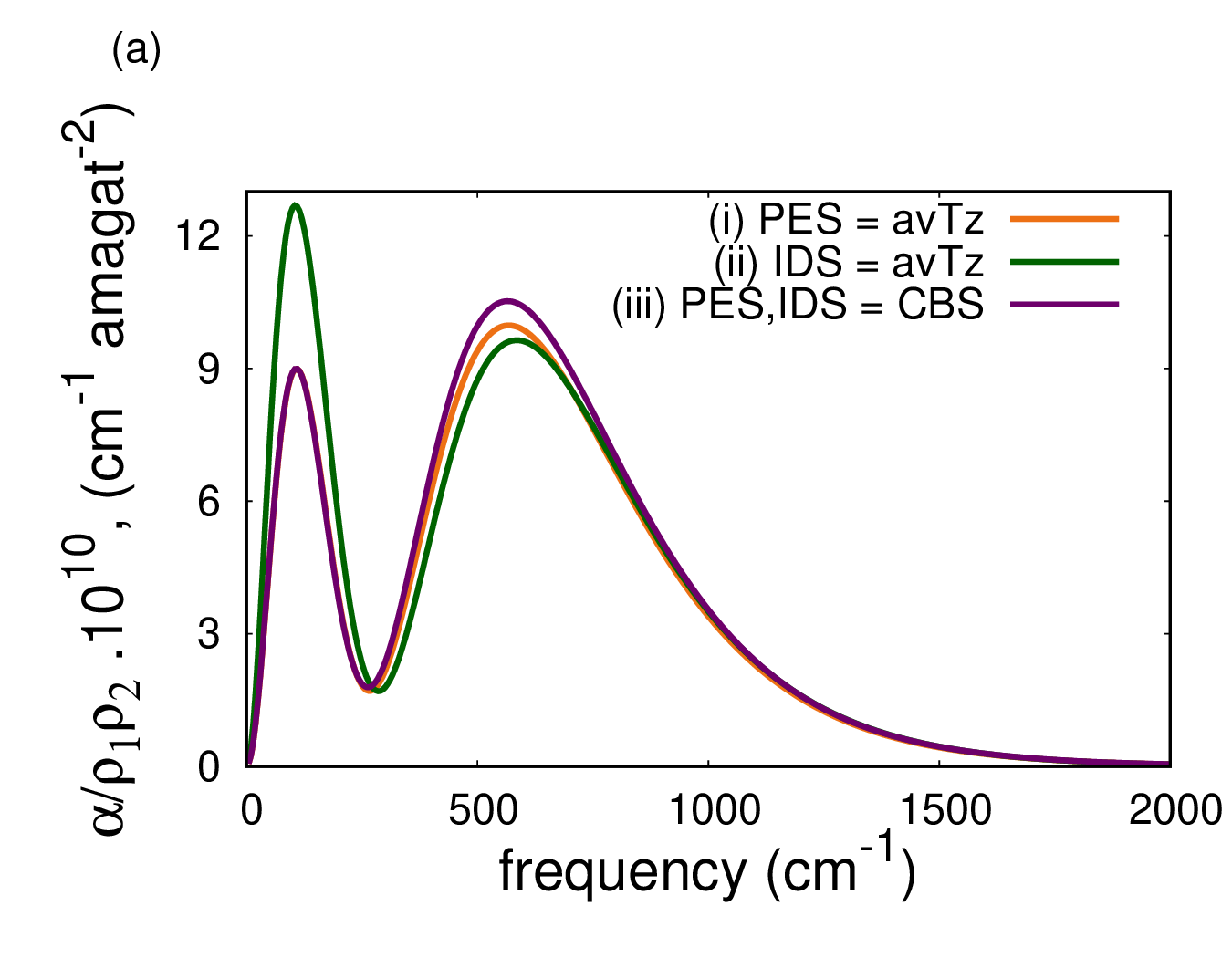}
         \includegraphics[width=0.475\textwidth]{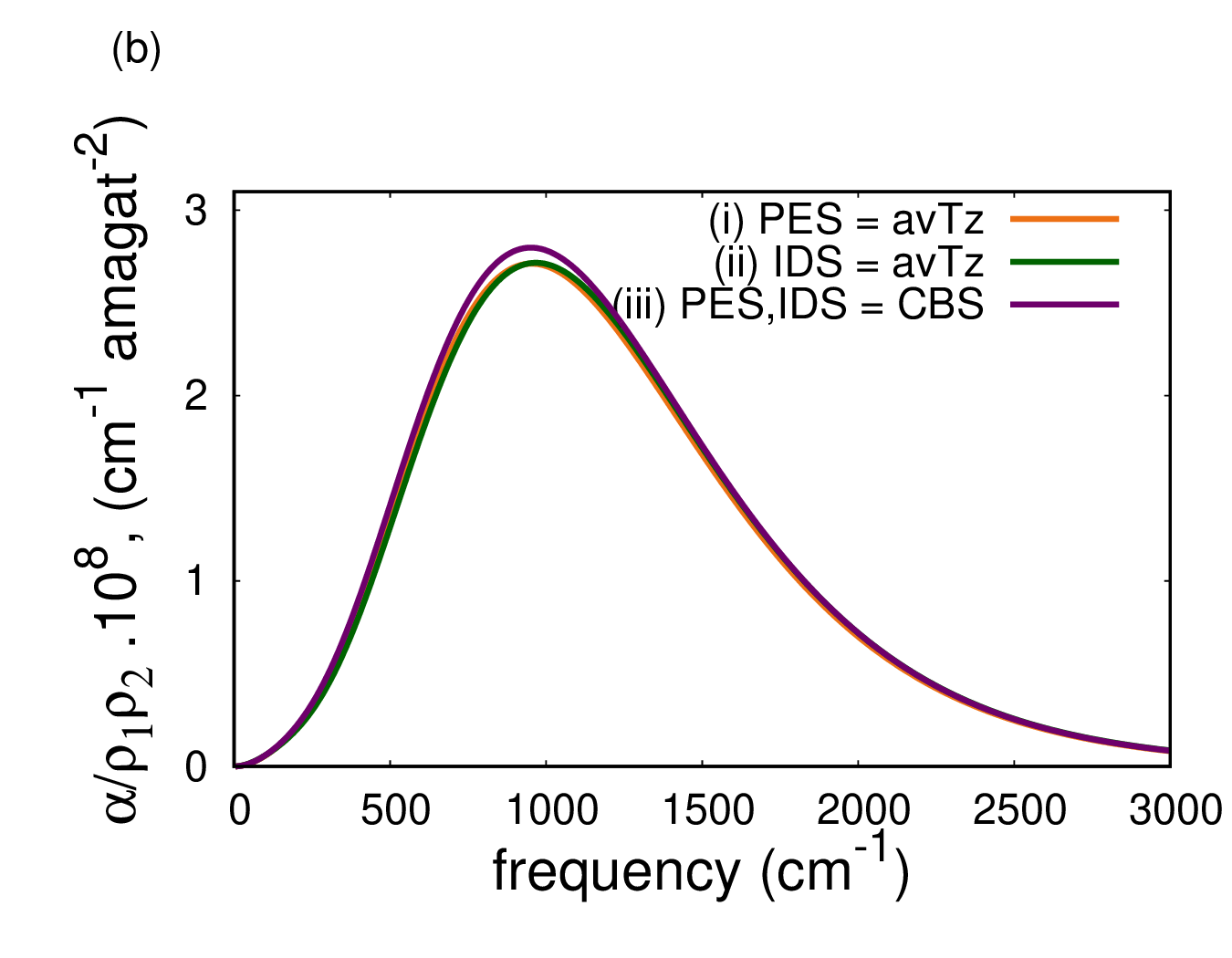}
         \includegraphics[width=0.475\textwidth]{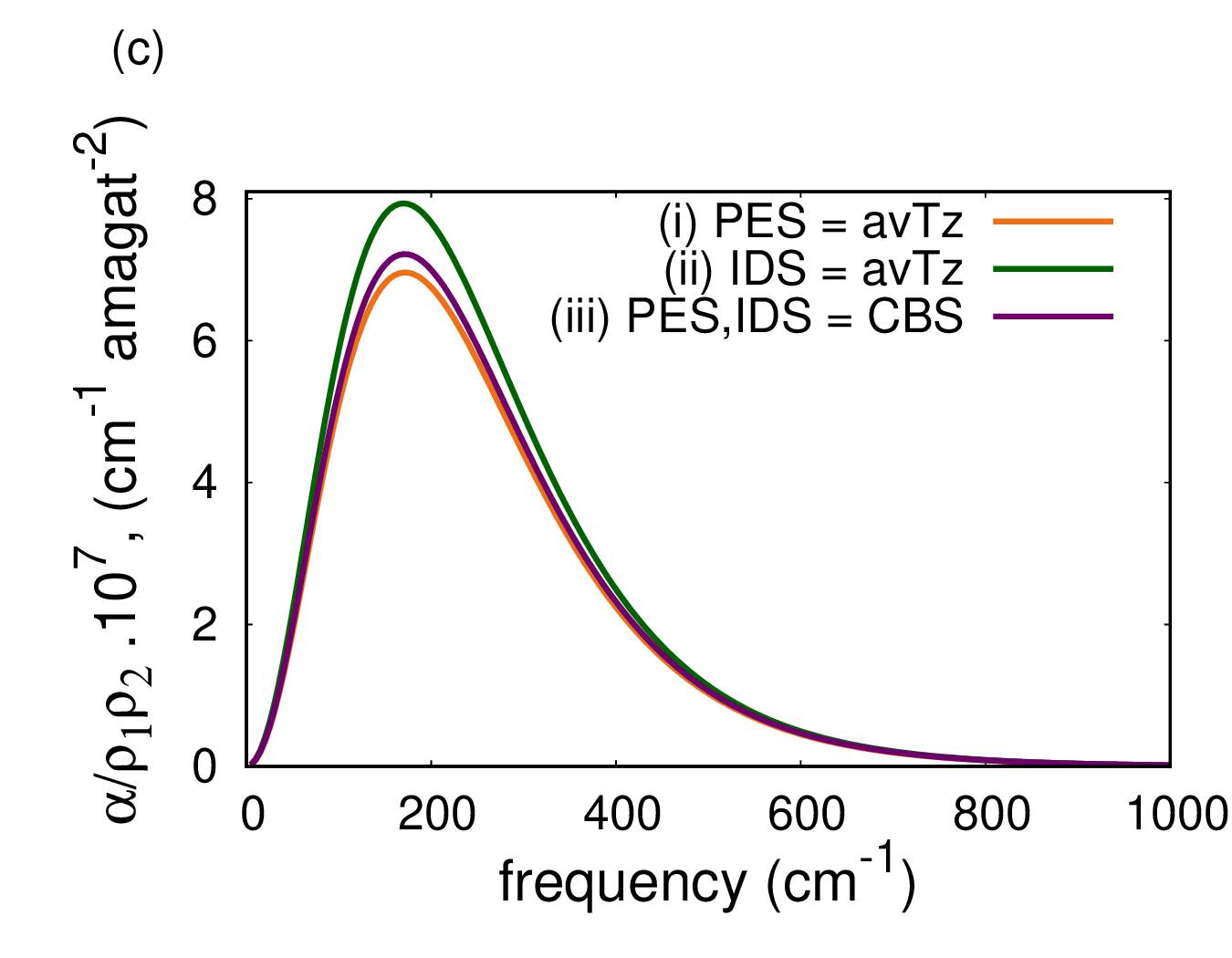}
         \includegraphics[width=0.475\textwidth]{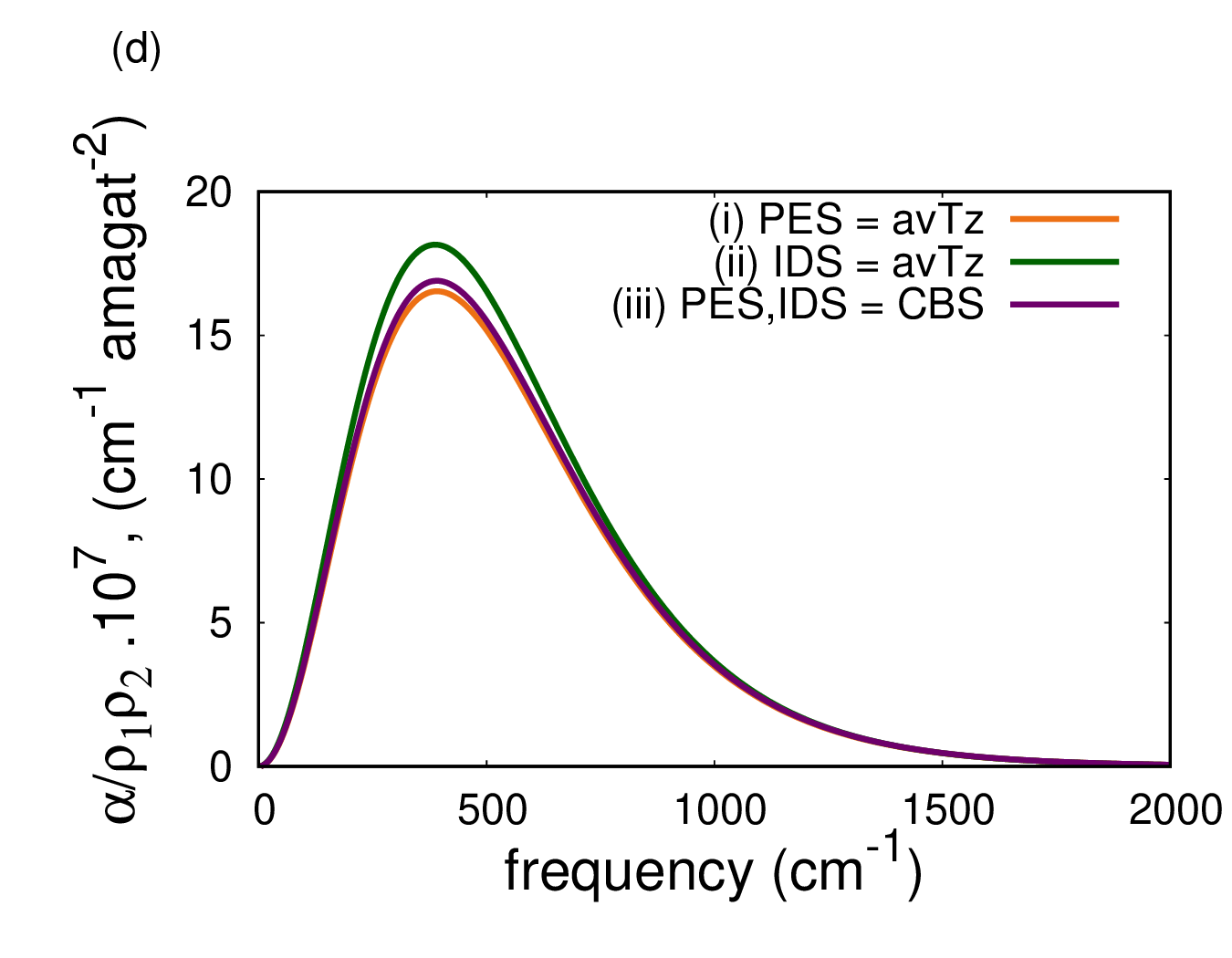}
         \includegraphics[width=0.475\textwidth]{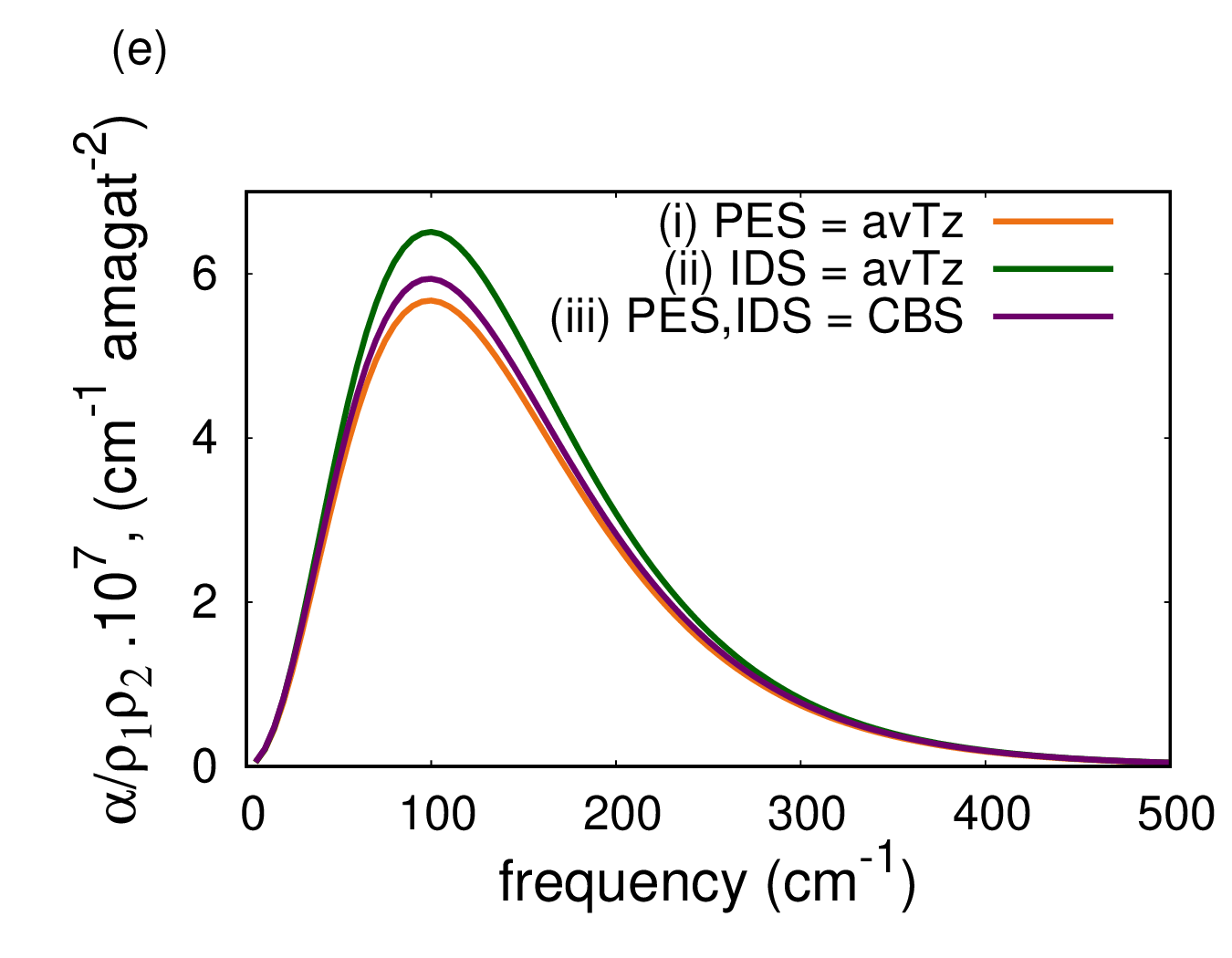}
         \includegraphics[width=0.475\textwidth]{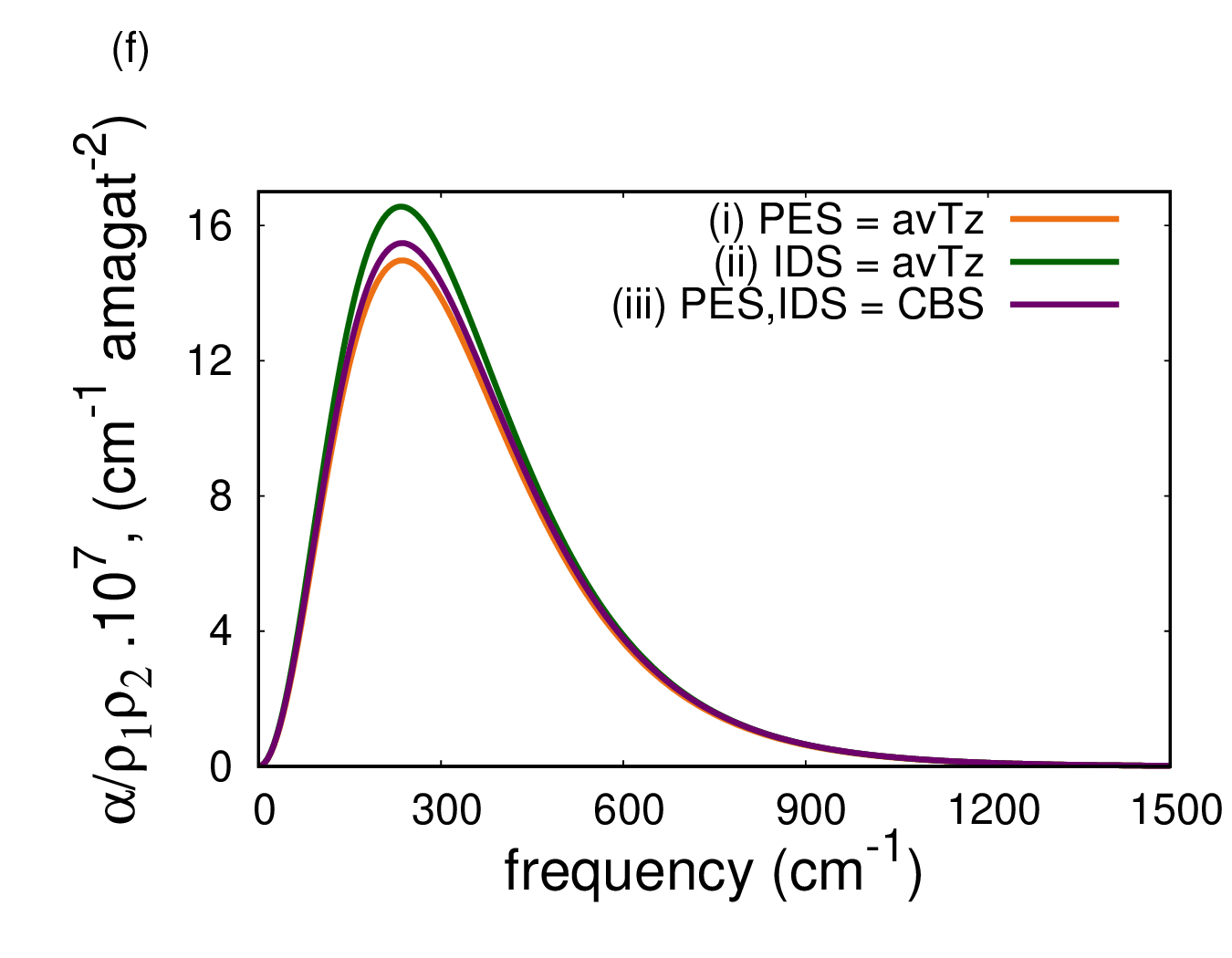}
         \caption{Comparison of CIA spectra computed with (iii) both the PES and IDS computed at at the CCSD(T)/CBS level, (i) with the IDS limited to the CCSD(T)/AVTZ level and (ii) with the PES limited to the CCSD(T)/AVTZ level, at two different temperatures. (a) and (b) show Ne--He at 295K and 2000K (c) and (d) show Ar--He at 295K and 2000K (e) and (f) show Ar--Ne at 295K and 2000K. }
         \label{diff_pes_ids}
\end{figure*}

We first investigate whether the spectrum is more sensitive to the PES, to the IDS, or similarly sensitive to both. 
We calculated spectra computed using our most accurate CCSD(T)/CBS level PES and IDS,
and repeated these calculations replacing either the PES or the IDS with the one computed at the CCSD(T)/AVTZ, \emph{i.e} computed in the smallest basis.
The comparison at two different temperatures is shown in Fig.~\ref{diff_pes_ids}.
For all systems at 295 K, the change in PES resulted in a difference of around 4\%, but the change in IDS caused a difference of around 10\% at the frequency with maximum intensity, see Table~\ref{tab:diff_295K}. These differences decrease at high temperatures, see Table~\ref{tab:diff_2000K}. 
Change in the dipole led to slight shifts in the spectra, which do not occur for similar changes in the potential. 
These results highlight the dominance of the dipole in determining the spectral outcomes, emphasizing its importance in accurate spectral predictions. 

\begin{figure*} 
         \centering
         \includegraphics[width=0.475\textwidth]{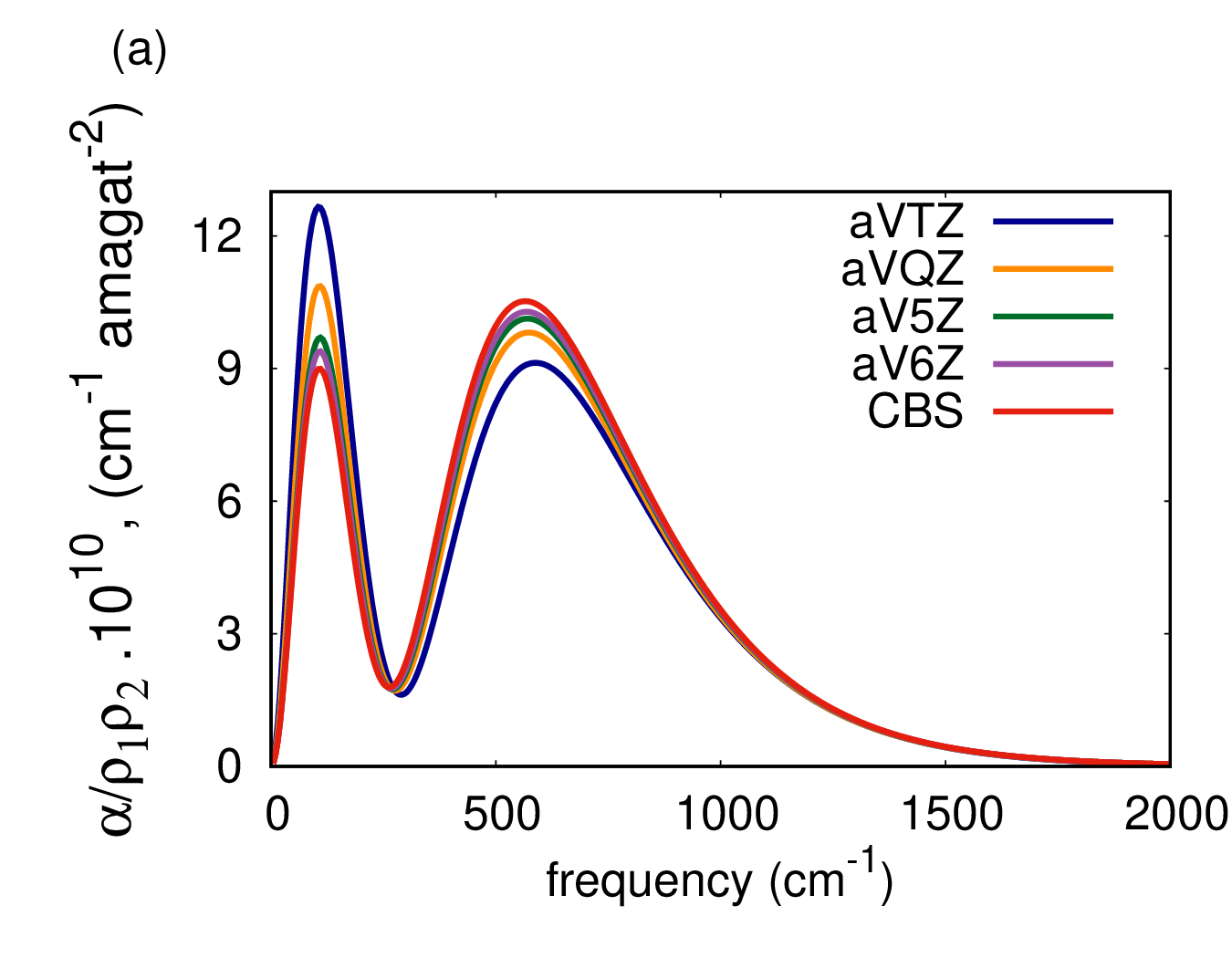}
         \includegraphics[width=0.475\textwidth]{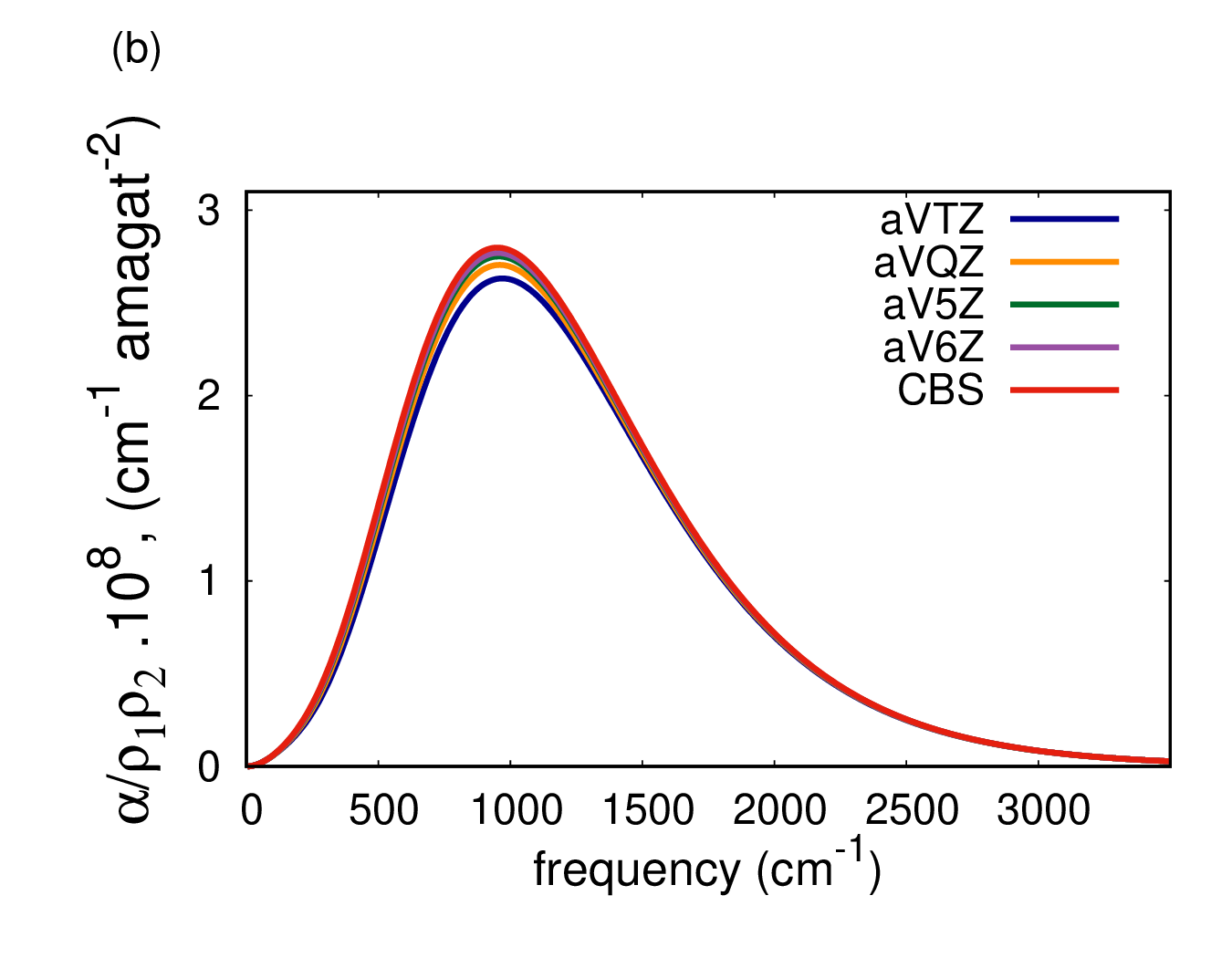}
         \includegraphics[width=0.475\textwidth]{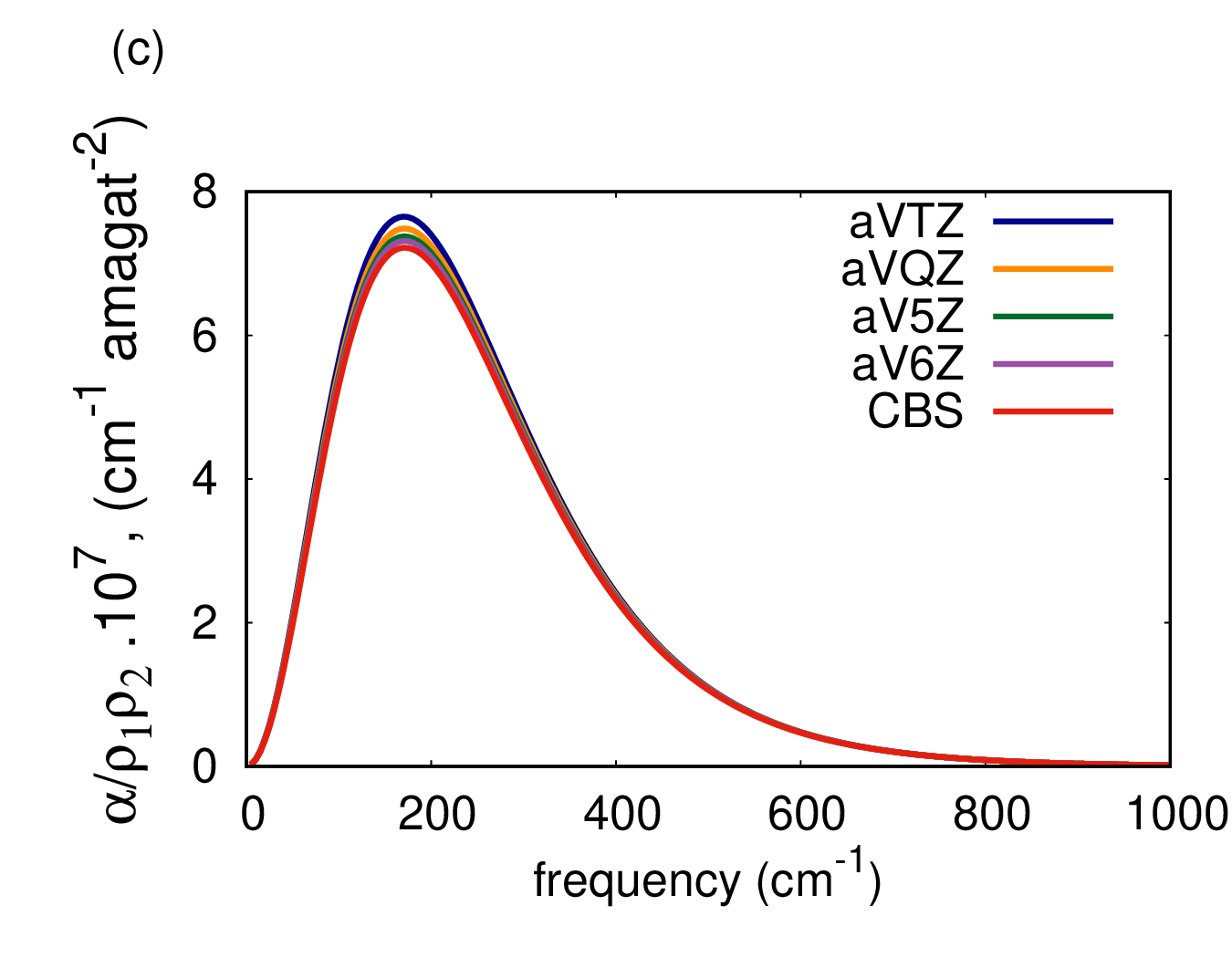}
         \includegraphics[width=0.475\textwidth]{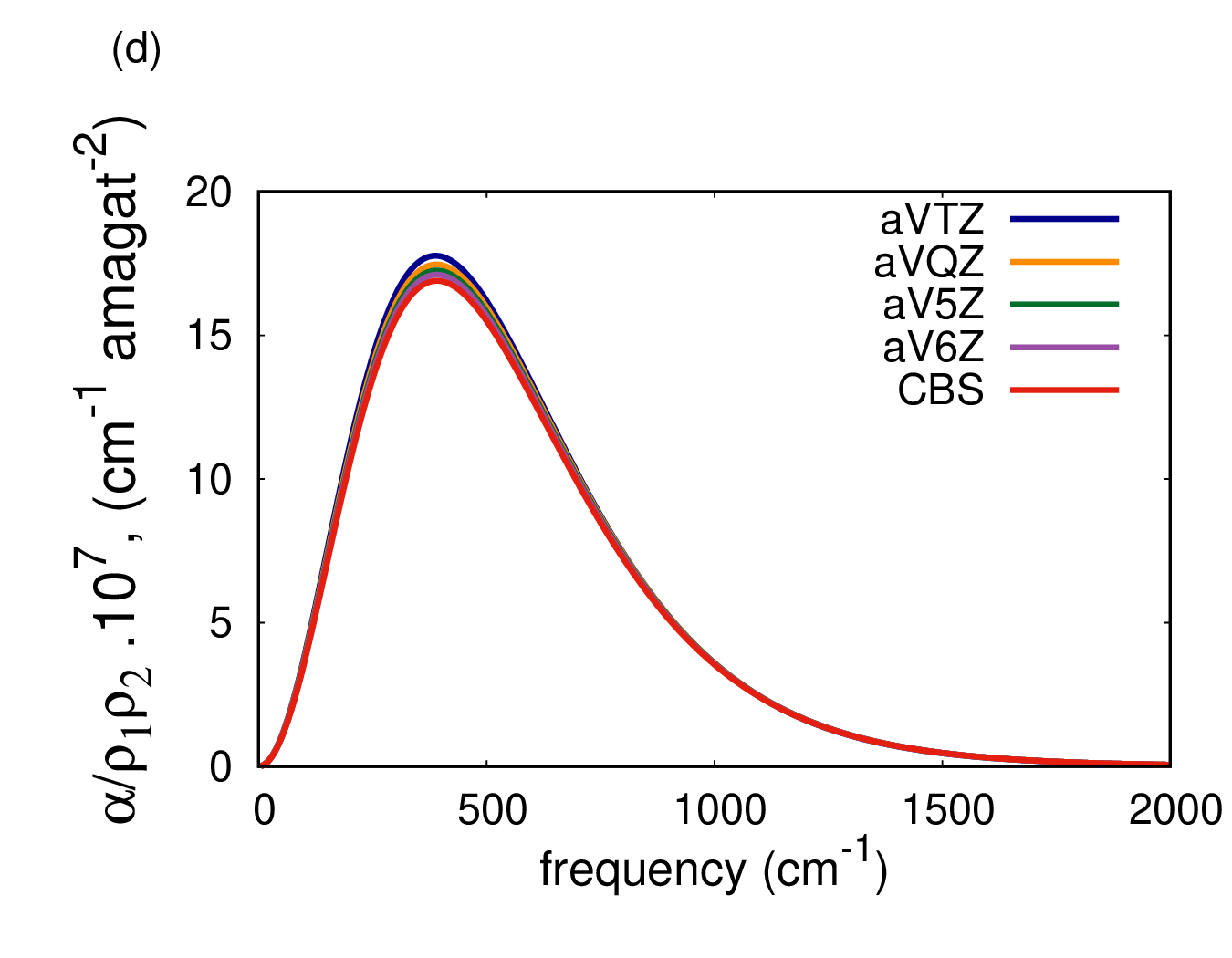}
         \includegraphics[width=0.475\textwidth]{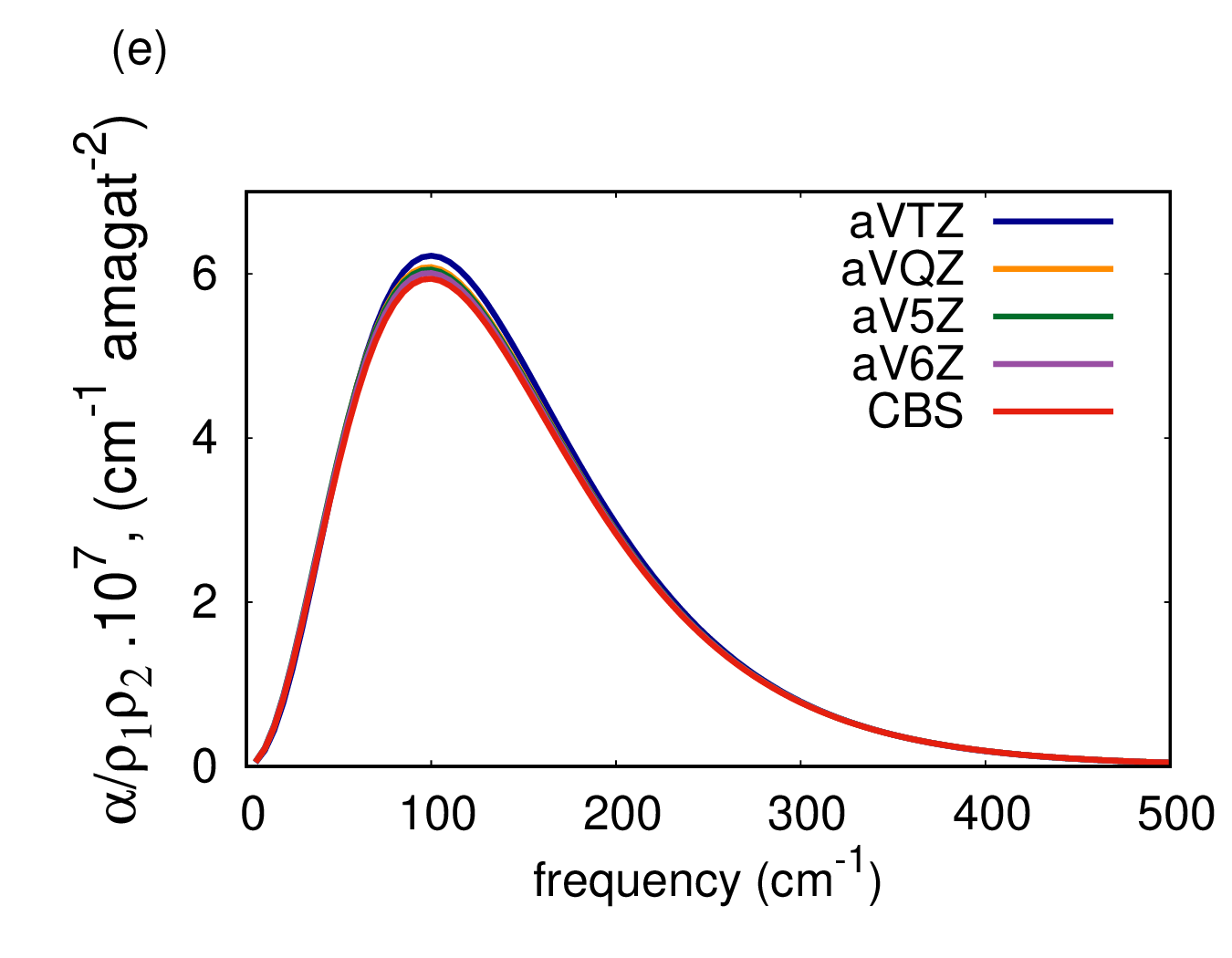}
         \includegraphics[width=0.475\textwidth]{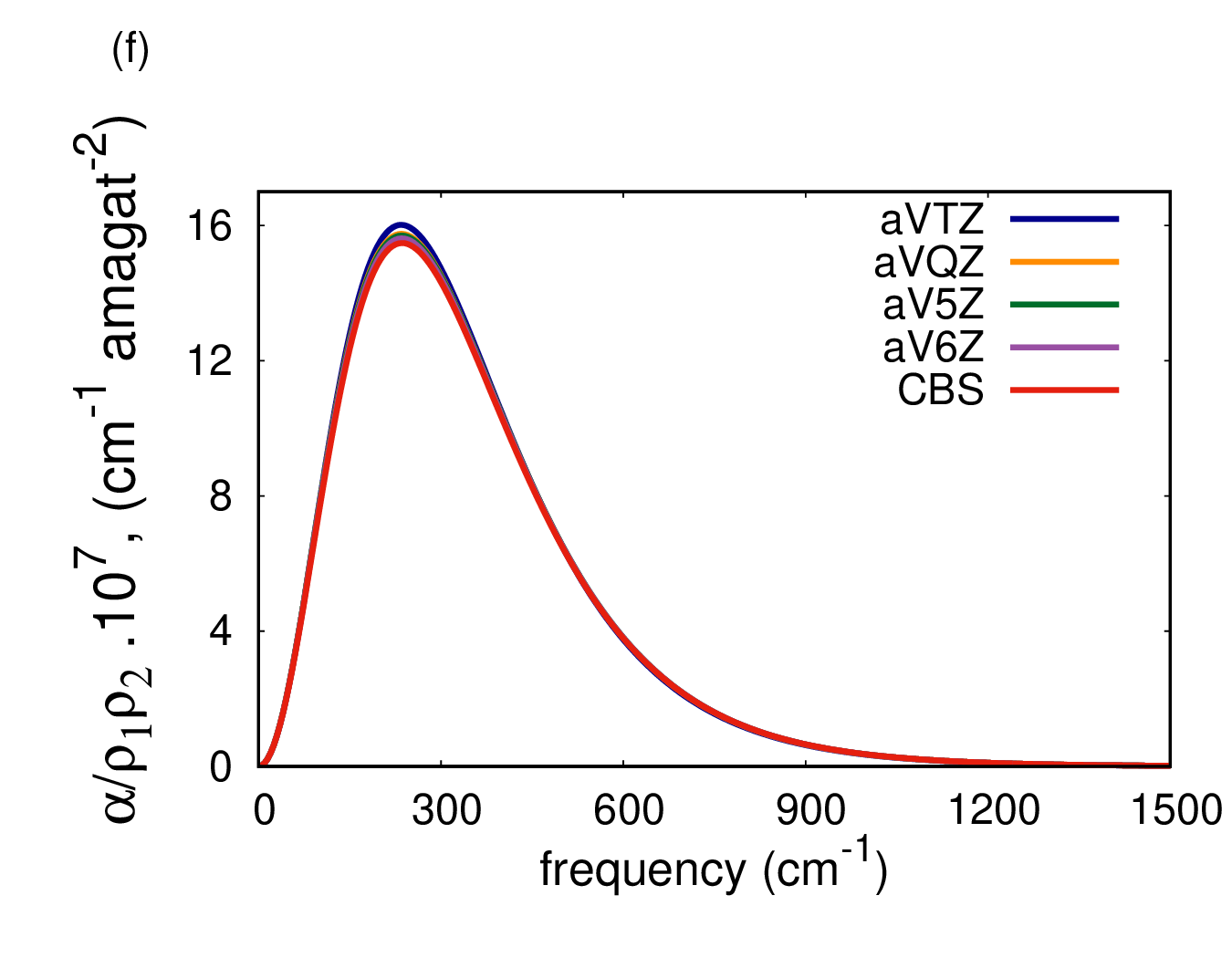}
         \caption{Convergence of CIA with different basis sets used in CCSD(T) calculations. (a) and (b) show Ne--He at 295K and 2000K (c) and (d) show Ar--He at 295K and 2000K (e) and (f) show Ar--Ne at 295K and 2000K }
         \label{CIA_convergence}
\end{figure*}

Next we consider the convergence of the CIA spectra as a function of the one-electron basis set, shown in Fig.~\ref{CIA_convergence}.
Here, spectra are computed using the PES and IDS computed in the indicated basis set.
The convergence is smooth and systematically goes to the spectra calculated with the highest basis set (CBS). To quantify the basis set dependence, we computed the relative difference between spectra calculated with different basis sets with the spectra calculated using the CBS. The results for each system at 295 K is shown in Table~\ref{tab:basis_295},
where the maximum error in the smallest basis set is around 5~\%.
At high temperatures, the convergence becomes even faster, see Table.~\ref{tab:basis_2000}.

\begin{figure*} 
         \centering
         \includegraphics[width=0.475\textwidth]{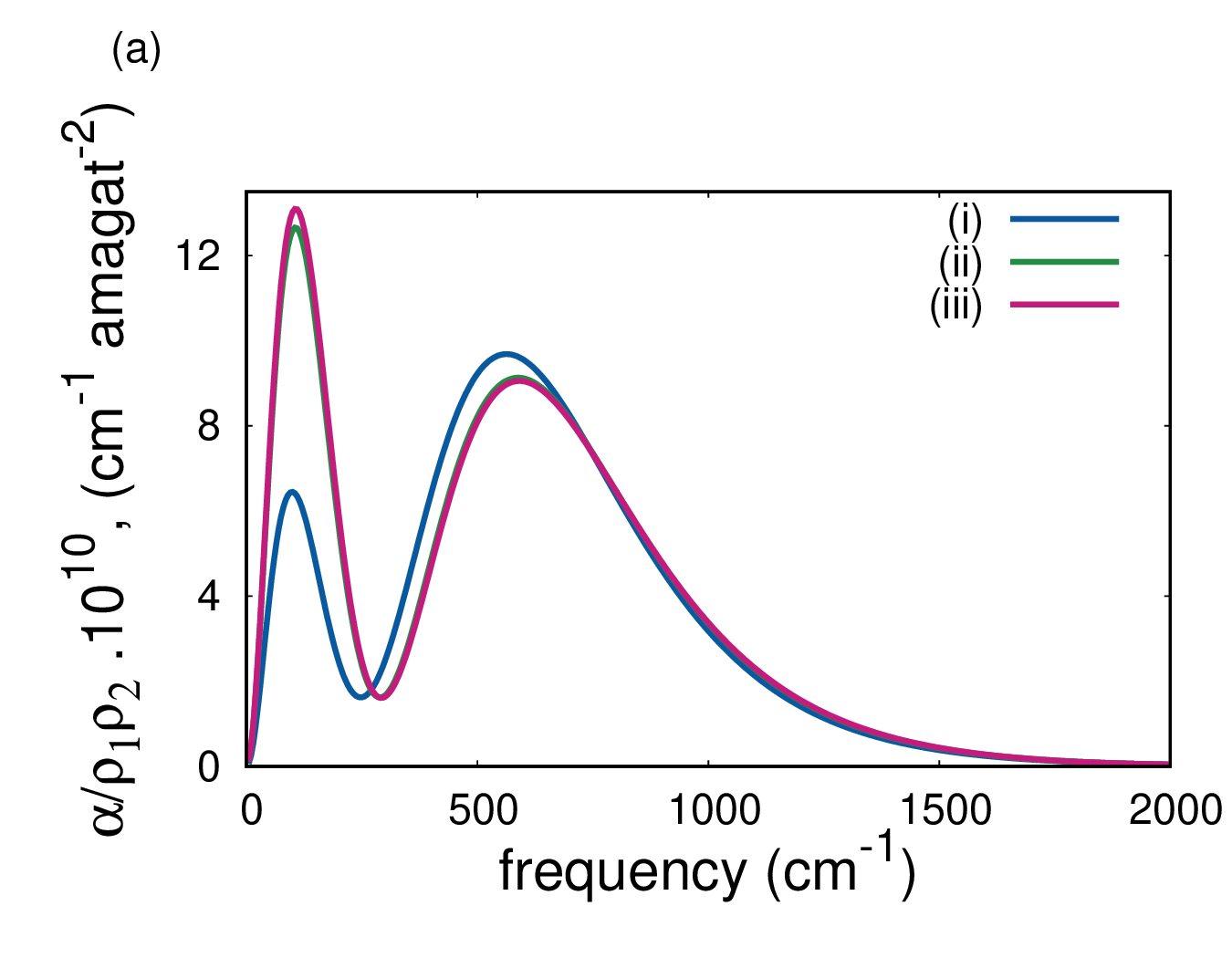}
         \includegraphics[width=0.475\textwidth]{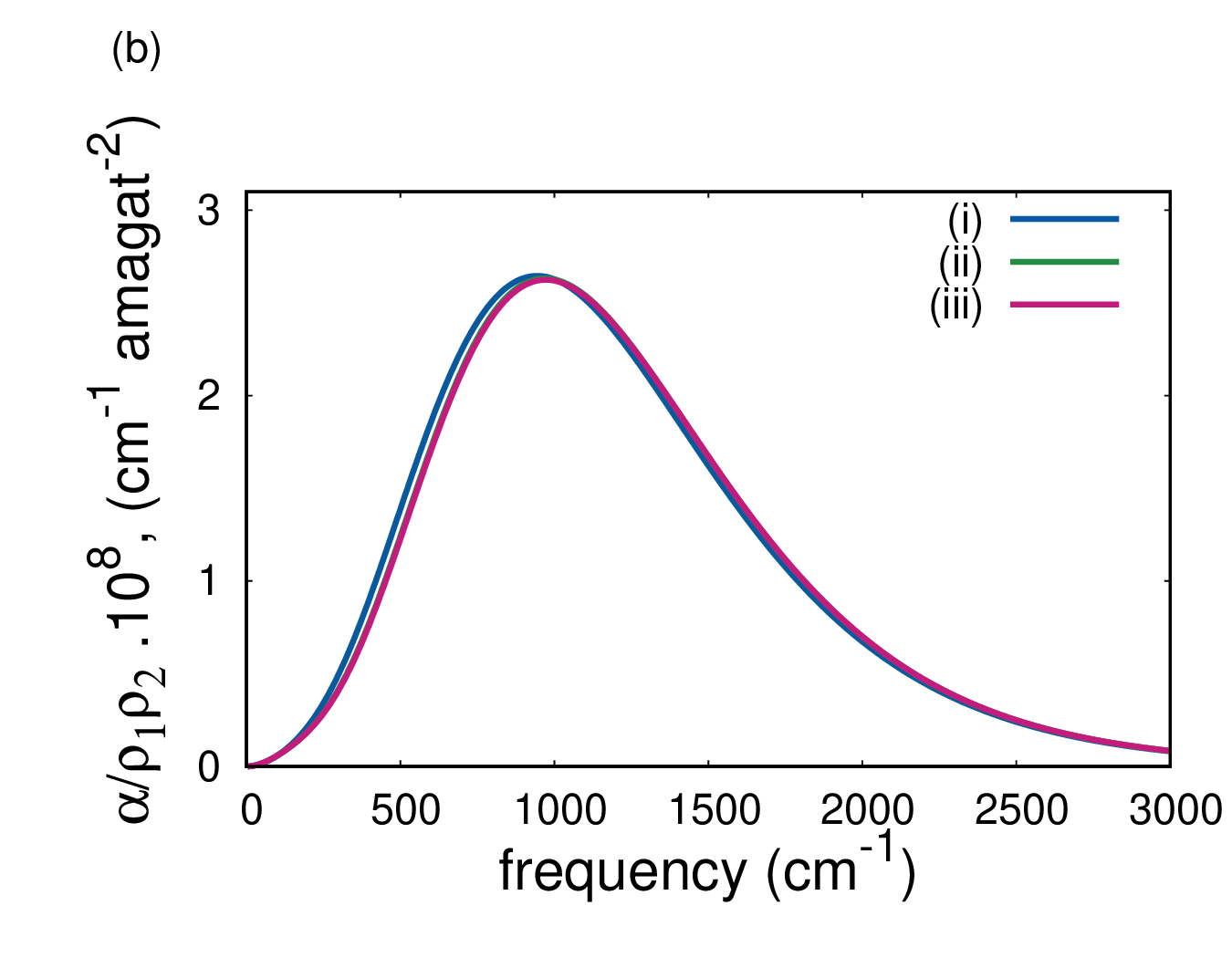}
         \includegraphics[width=0.475\textwidth]{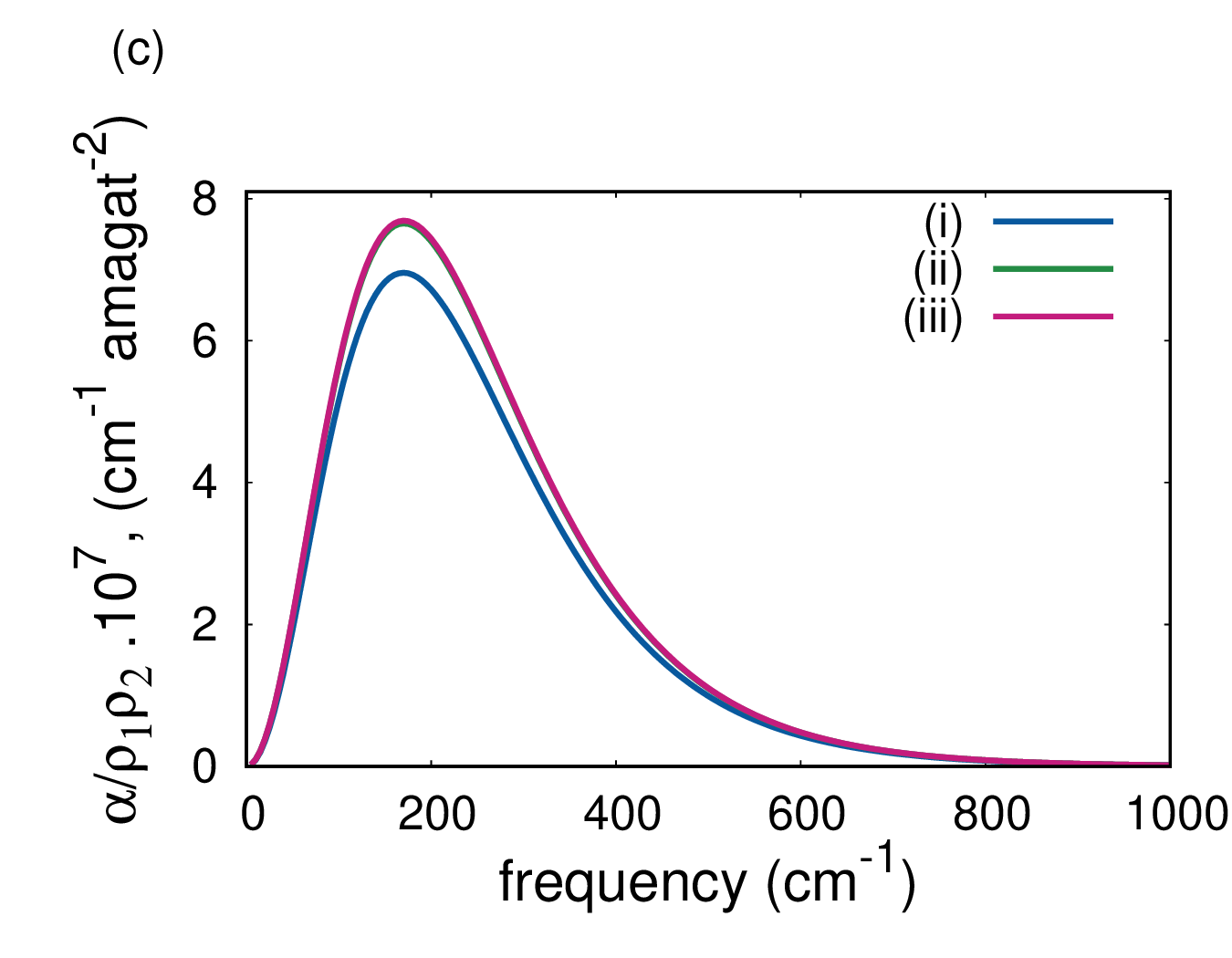}
         \includegraphics[width=0.475\textwidth]{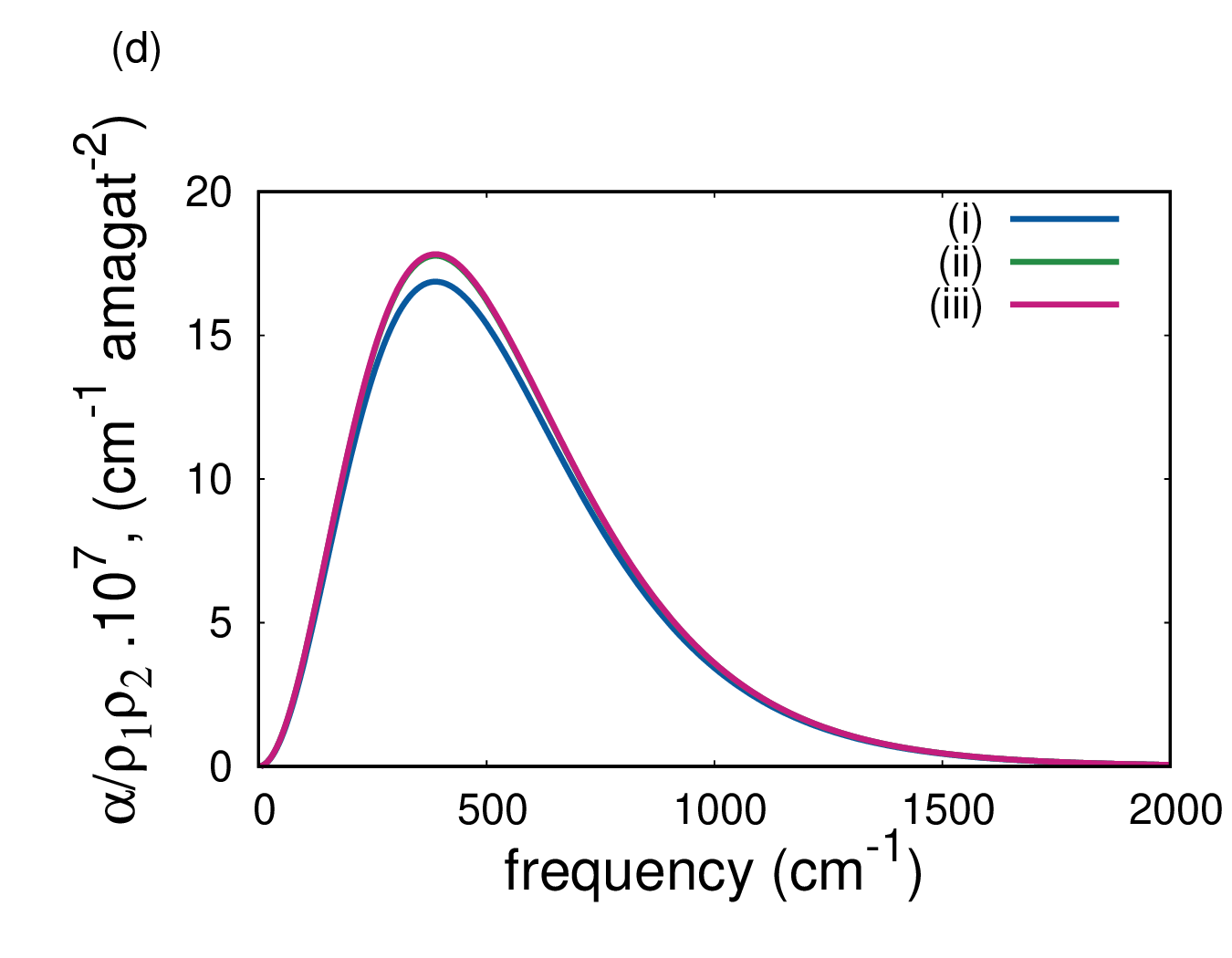}
         \includegraphics[width=0.475\textwidth]{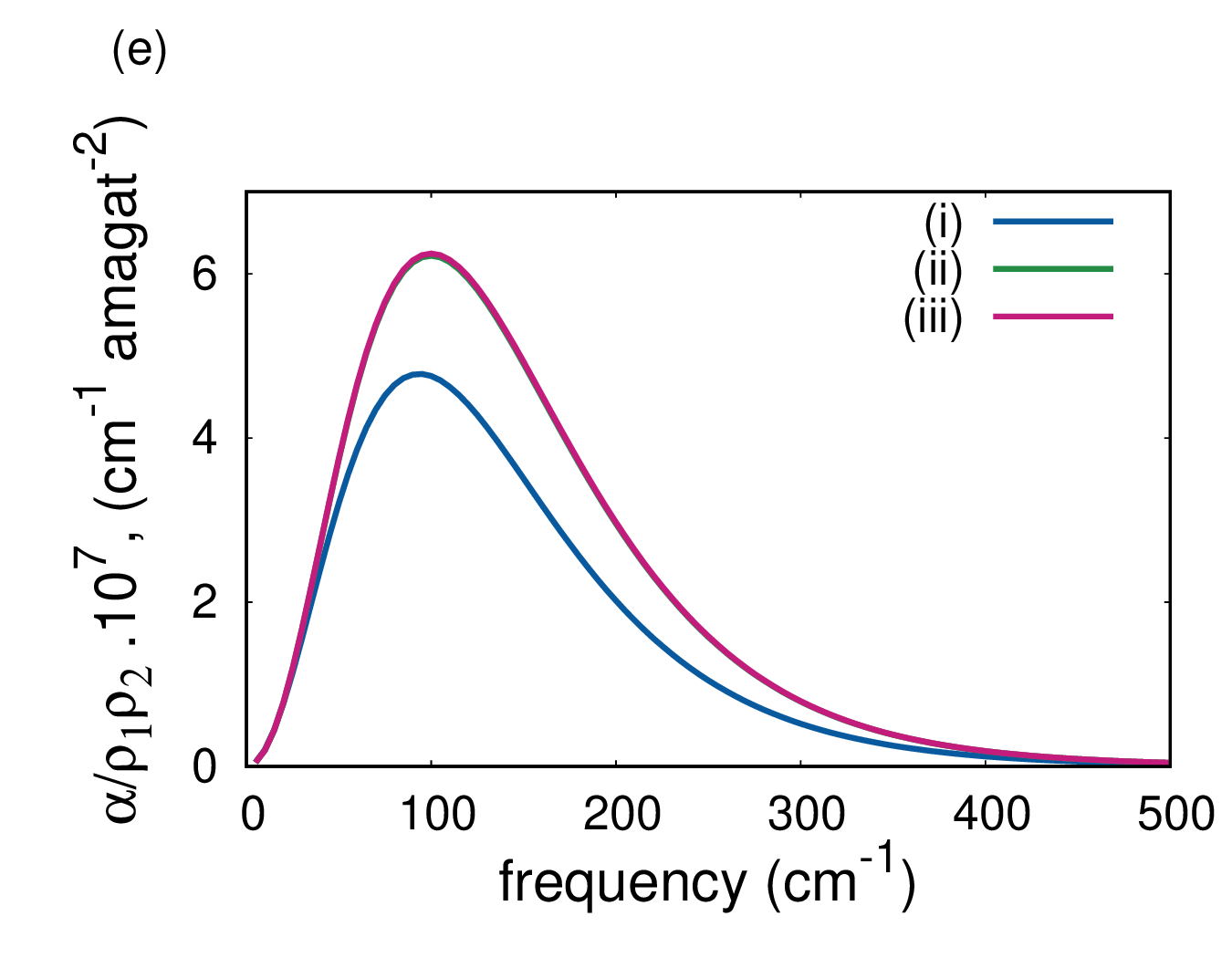}
         \includegraphics[width=0.475\textwidth]{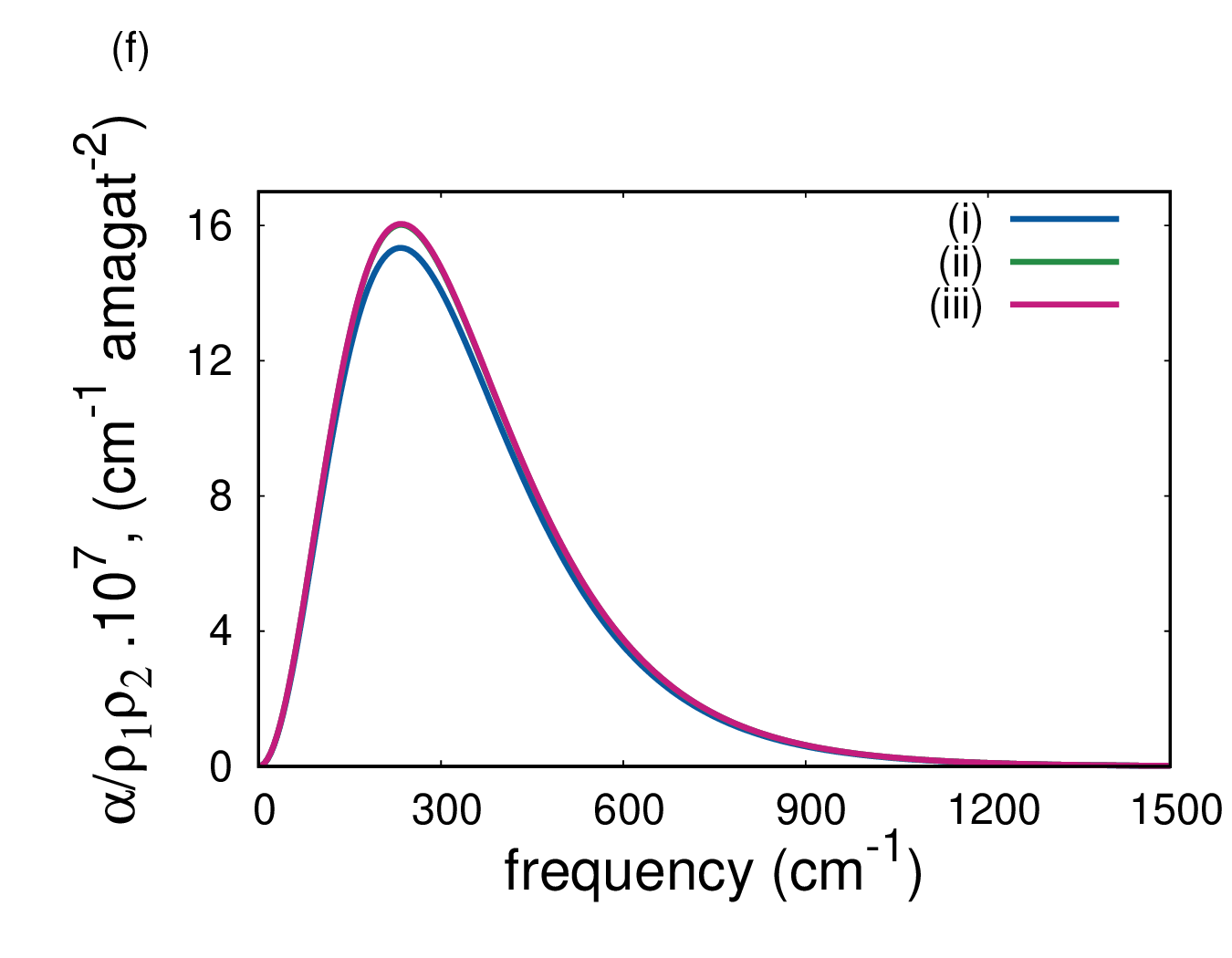}
         \caption{Comparison of CIA calculated using different methods. All calculations employed the aug-cc-pVTZ (AVTZ) basis set supplemented with midbond functions. (i) CCSD/AVTZ (ii) CCSD(T)/AVTZ (iii) FCI extrapolation (a) and (b) show Ne--He at 295K and 2000K; (c) and (d) show Ar--He at 295K and 2000K (e) and (f) show Ar--Ne at 295K and 2000K.}
         \label{methods}
\end{figure*}

To evaluate the influence of the level of theory on computed spectra, we calculated spectra using CCSD, CCSD(T), and an extrapolation to the full configuration interaction (FCI) limit based on the continued fraction extrapolation scheme, Eq.~(6) of \cite{fci}. All calculations employed the aug-cc-pVTZ (AVTZ) basis set supplemented with midbond functions. 
The resulting spectra are compared in Fig.~\ref{methods}.
A substantial difference is observed between the CCSD and CCSD(T) results, highlighting the significant role of triple excitations in recovering the correlation energy.
In contrast, the spectra obtained from CCSD(T) and the FCI show minimal deviation, indicating that the coupled cluster series rapidly converges. CCSD(T) captures majority of the dynamic correlation relevant to the spectral features, see Tables~\ref{tab:method_295}, and \ref{tab:method_2000}. 
The difference between the CCSD calculations and the estimated FCI limit is in the order of 10~\% at 295~K.

We use the integrated intensity to quantify the difference between the spectra, which mostly agree in overall shape.
Using a sum rule evaluated in a classical statistical mechanical approximation, the integrated intensity is readily evaluated numerically at a given temperature as an integral over the square dipole moment and Boltzmann factor of the potential, see Eq.~\eqref{csm}.
We evaluated the relative difference in the integrated intensities between spectra calculated at the CCSD(T)/AVTZ and CCSD(T)/CBS, as well as CCSD/AVTZ and FCI/AVTZ, representing the worst and best cases in the convergence of basis set study and convergence of different methods study.
The latter relative difference was greater, constituting a very conservative estimate of the uncertainty in the spectra computed using our most accurate CCSD(T)/CBS potentials and dipole moments.
We then estimate a theoretical error bar for the spectrum, by applying an overall scaling by plus or minus this relative difference in the integrated intensity.

The integrand, $F(R)$ of Eq.~\eqref{csm} can be interpreted as the contribution from a specific intermolecular distance, $R$, to the integrated intensity.
We use this here to quantify the range of intermolecular distances the CIA spectra are sensitive to.
Figure~\ref{range} shows the contribution of various values of $R$ for all systems studied for various temperatures.
The spectrum is typically determined by a range of $R$ of about a bohr wide,
determined by the balance between the decay of $D(R)$ for larger $R$,
and the suppression of $\exp(-\mathcal{V}/k_BT)$ at short distances where the potential becomes repulsive.
As the temperature increases, shorter distances become progressively more important.
However, at room temperature and even at lower temperatures down to 10~K,
we find that there is no contribution from the long-range region beyond 8~$a_0$ where the van der Waals interaction determines the dipole moment.
This is somewhat surprising in view of the effort invested in computing van der Waals contributions to the interaction induced dipole \cite{bohr1,bohr2,fowler,heijmen1996symmetry,meyer1994long,bishop1993calculation,fowler1990d9}.

\section{Impact of Dipole Surface on Spectral Features} 

While previous studies consistently reported single peak spectra for all systems (Ne--He, Ar--He, and Ar--Ne), our calculations reveal an unexpected feature in the Ne--He spectrum at some temperatures. 
Rather than following the anticipated smooth, single peak profile, as shown in \cite{hene_exp_2,hene_exp,hene_3,K_W,hene4,elkader}, Ne--He spectra exhibit a distinct dip.  

To better understand how the shape of IDS influences the spectral profile, we analyzed two representative dipole models. 
The first model dipole, commonly used in earlier studies, adopts a single exponential form.
The second model dipole is a double exponential which captures the zero crossing near 4.2 bohr.

To construct these dipole models we fit the log of the dipole moment with straight lines in two regions, as $\log B - \beta R$ between 2 and 3.6~$a_0$ and $\log A - \alpha R$ between 4.7 and 6~$a_0$.
The double exponential model is then given as $\mathrm{D(R)}=  A ~\mathrm{exp}(-\alpha \mathrm{x}) - \mathrm B~{\exp}(-\beta \mathrm{x})$,
while for the single-exponential model we keep only the longer ranged exponential that determines the asymptotic decay.

\begin{figure*} 
         \centering
         \includegraphics[width=0.475\textwidth]{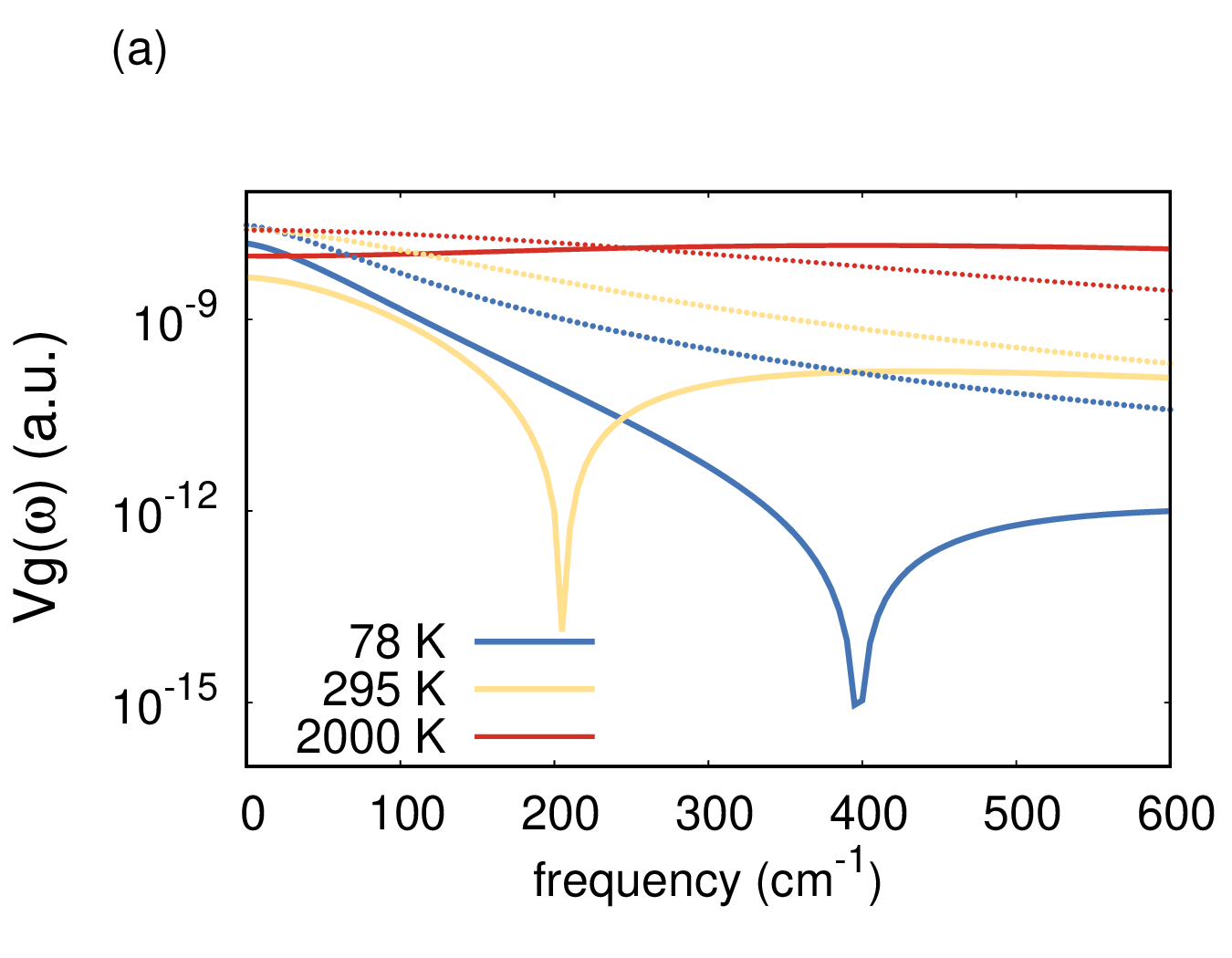}
         \includegraphics[width=0.475\textwidth]{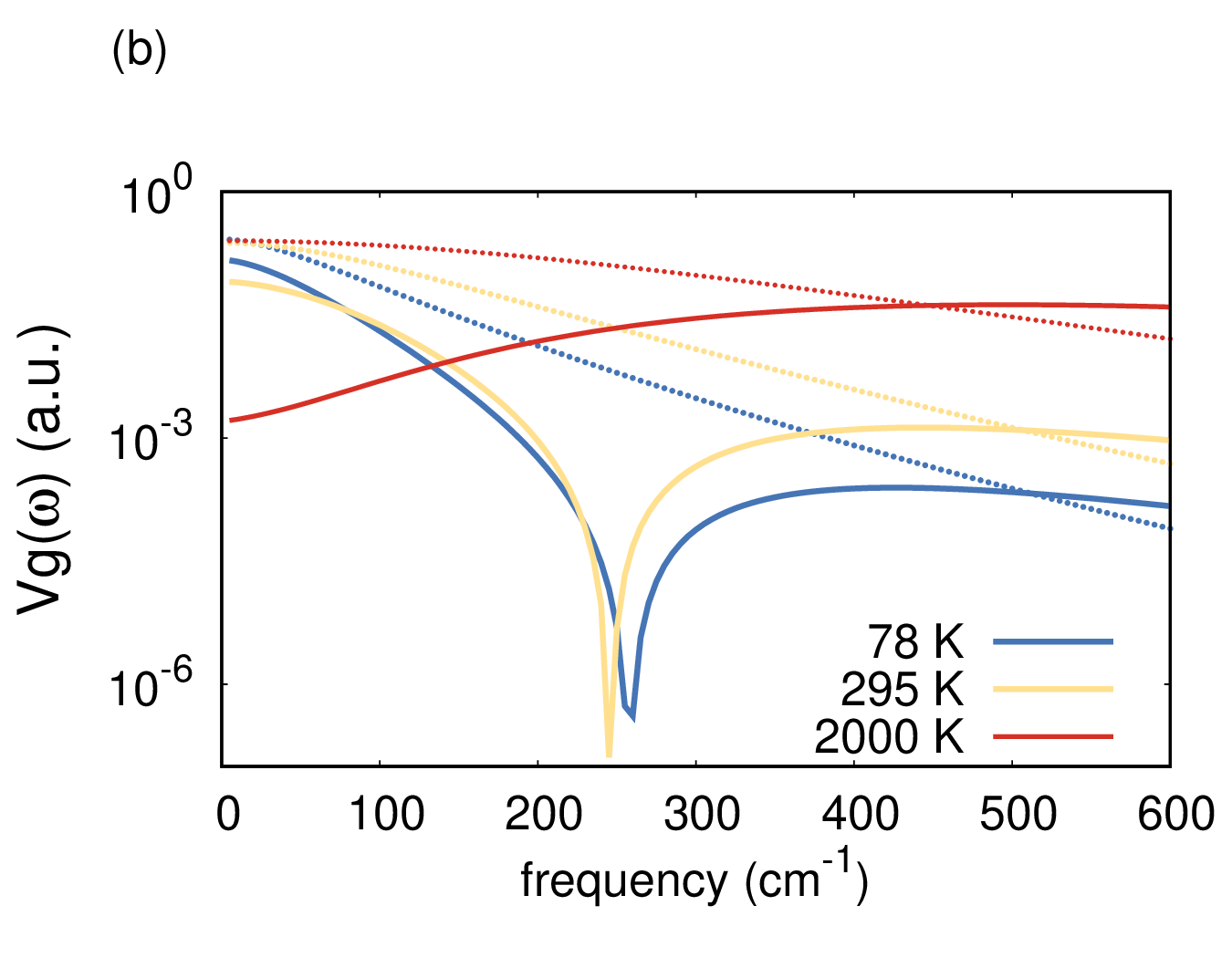}

         \caption{Approximate lineshapes in the (a) hard-sphere approximation and (b) soft-sphere approximation. Dotted (solid) lines show results for the single (double) exponential dipole model.}
         \label{diff_fou}
\end{figure*}

We then used the single and double-exponential dipole models to compute the lineshape in the hard-sphere approximation.
Figure~\ref{diff_fou} shows that the double-exponential dipole gives rise to additional features that are absent in single-exponential dipole model.
In the hard-sphere approximation, the lineshape is related to the Fourier transform of the dipole function, which for a single exponential is a Lorentzian.
The Lorentzian decays monotonically and cannot describe a local minimum at intermediate wavenumbers.
The double-exponential dipole, by contrast, crosses zero close to $R= 4.2$~$a_0$,
and in the vicinity of this point the dipole functions appears to oscillate with a certain wavelength, that is associated with a certain frequency that becomes visible in its Fourier transform.

\begin{figure*} 
         \centering
         \includegraphics[width=0.475\textwidth]{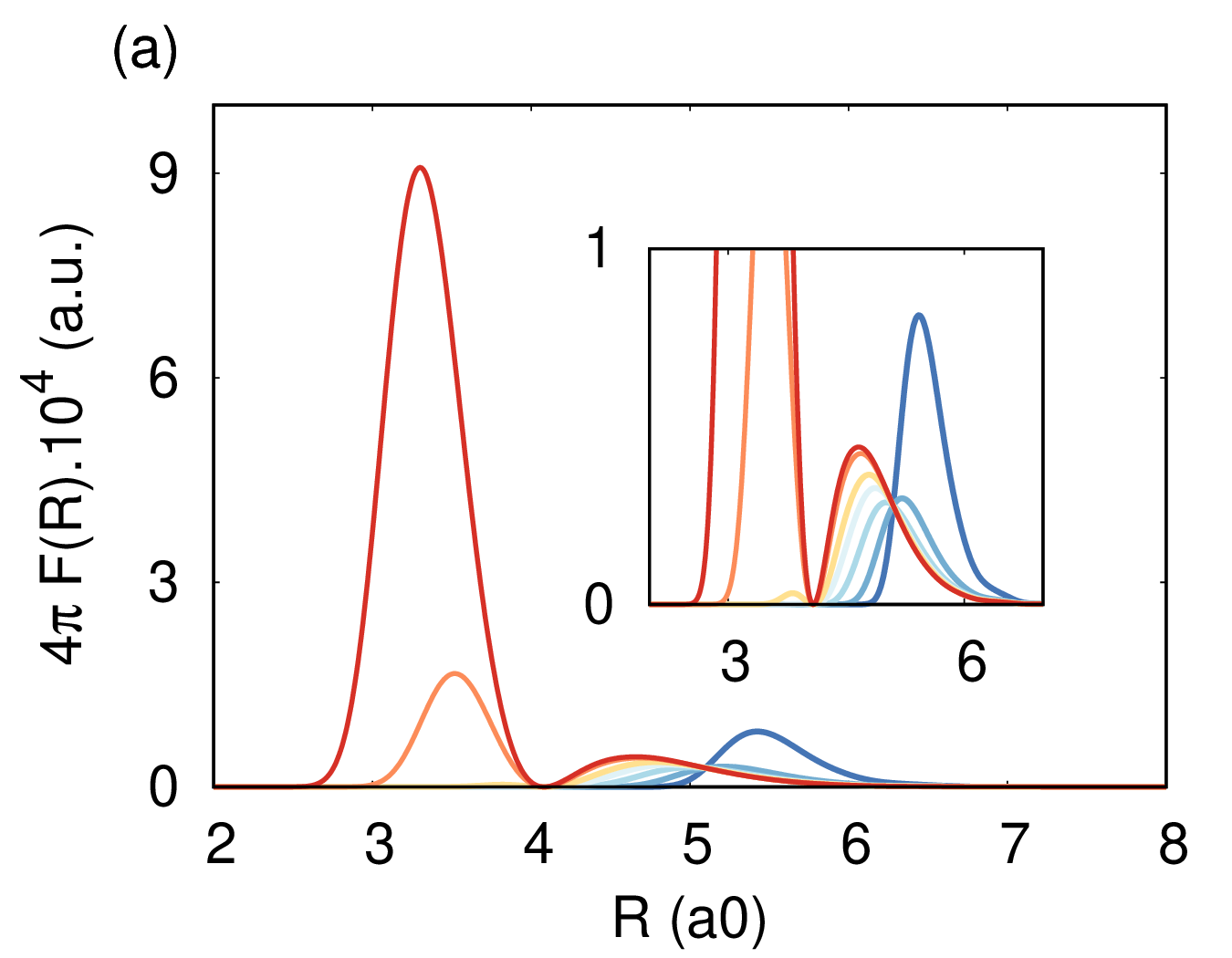}
         \includegraphics[width=0.475\textwidth]{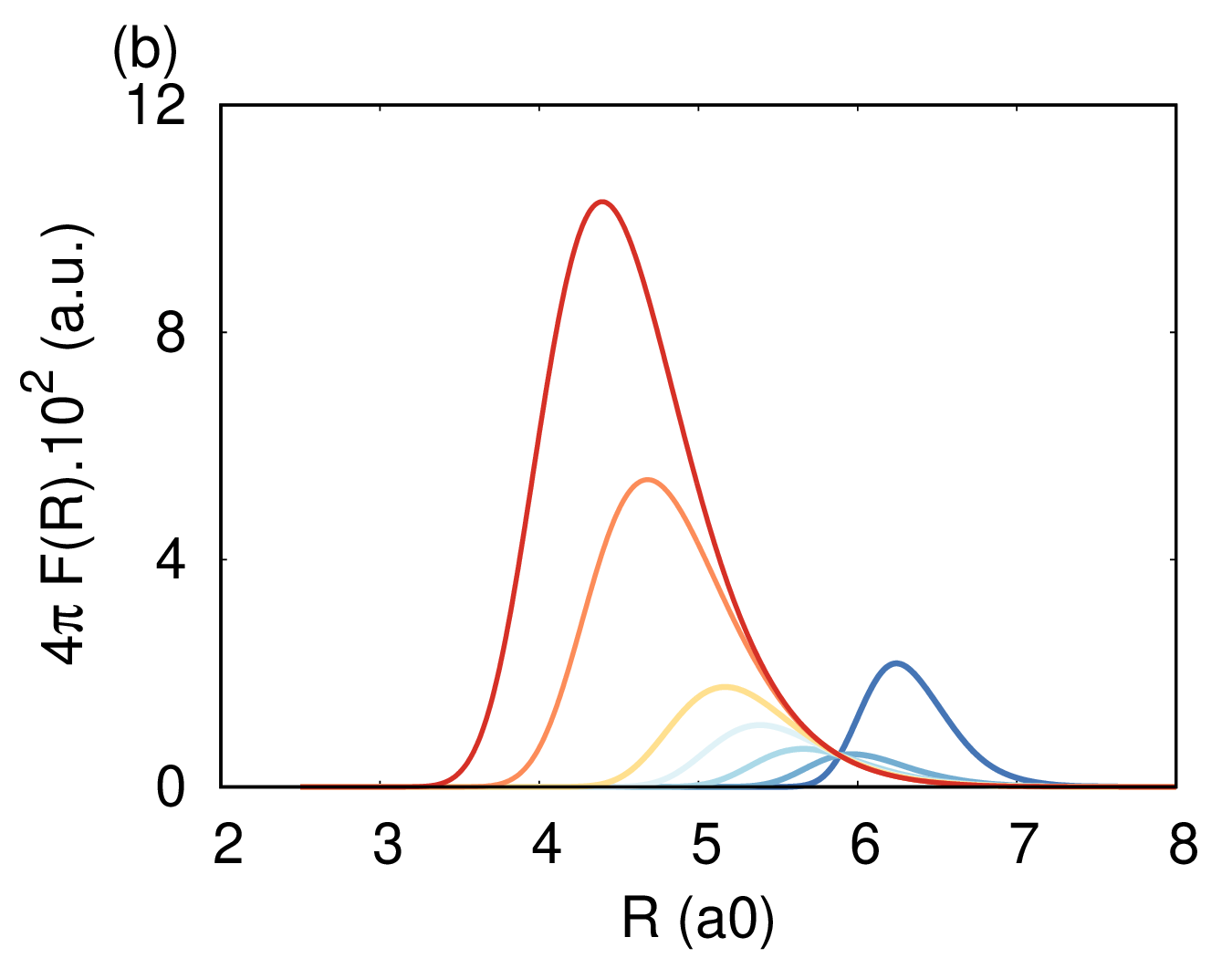}
         \includegraphics[width=0.475\textwidth]{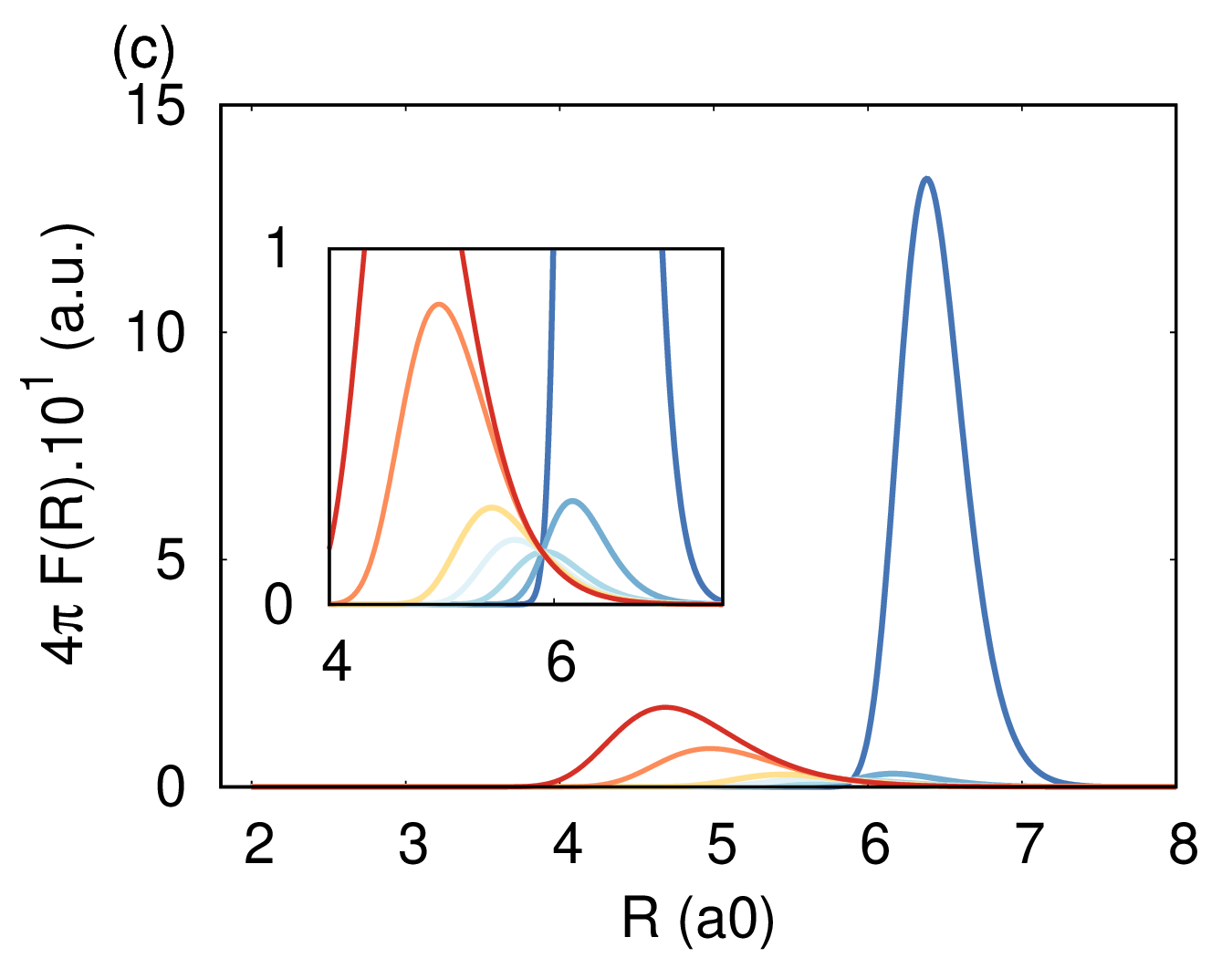}
         \includegraphics[width=0.475\textwidth]{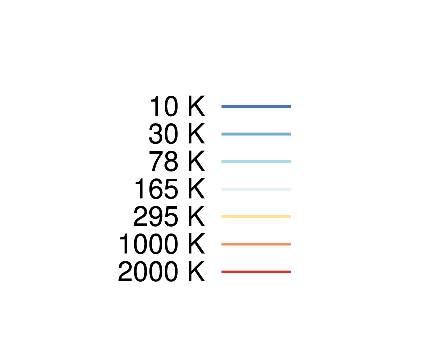}
         \caption{Contribution of intermolecular distances $R$ to the integrated intensity at different temperatures. Different panels correspond to (a) Ne--He (b) Ar--He (c) Ar--Ne.}
         \label{range}
\end{figure*}

Another way to think about the failure of single-exponential dipole models may be the following.
The spectrum is dominated by a narrow range of distances, $R$, see Fig.~\ref{range}.
Thus one could think of computing the spectrum using a dipole model that does not necessarily provide a good fit of the dipole function globally, but does fit reasonably in the limited range that matters.
If this dipole model is chosen to be a single monotonically decaying exponential, it cannot describe certain features such as zero crossings or local maxima in the dipole function.
By contrast, for an accurate dipole function;
if, as a function of temperature, such features enter the range of distances that determine the absorption spectrum, see Fig.~\ref{range}, corresponding spectral features can occur.
For example, for Ne--He, we predict that double peaked structure is most prominent around room temperature,
and disappears at both the lower and much higher temperatures of 77~K and 2000~K, respectively.

We also computed the lineshape in the soft-sphere approximation, see Fig.~\ref{diff_fou}(b).
Accounting in this way for the interaction potential impacts the results only at low temperatures, and affects only the position of the dip in the spectrum, not whether or not it occurs.
Thus, in both the hard and soft-sphere approximations, the dipole function with a zero crossing consistently leads to additional spectral structure in the transformation of the dipole. In contrast, the single-exponential dipole model fails to capture these features. 

These findings emphasize that an oversimplified short-range dipole model, such as those based on single exponential forms --which are quite common\cite{hene_3,elkader}-- may fail to reproduce important spectral details.

\section{Comparison to experiment}

Our results are compared with the experimental spectra of Refs.~\cite{CIA_exp,hene_exp, CIA_exp_bw1,bt1,bt2,bt3}. For both Ar--Ne and Ar--He, the calculated spectra are in good agreement with the experimental data in terms of both the spectral shape and the intensity. However, the experimental spectra show slightly higher intensities than the calculated ones, with a difference of about 9\% at 165 K and reduces to around 5\% at 295 K. The experimental results for Ar--He and Ar--Ne agree with our calculations to within their conservatively estimated uncertainty, whereas no experimental bars were reported. In contrast, the spectrum of Ne--He measured at 77 K \cite{hene_exp} is twice as intense as the calculated one and does not agree with our calculations to even within its conservatively estimated uncertainty.
The experimental data points were recorded only up to 250~cm$^{-1}$,
which coincides with the minimum in the calculated spectrum.
At a higher temperature of 295~K, this minimum should be more pronounced, but we are unaware of experimental results that test our prediction.
References~\cite{bt1,K_W,levinequantum,hene_exp_2} discuss CIA in Ne--He mixtures at 295~K,
but report data only for three wavenumbers 200, 400, 600~cm$^{-1}$.
Moreover, the reported absorption coefficients disagree with each other by up to a factor of four, and disagree with our calculations by two orders of magnitude.

\begin{figure*} 
         \centering
         \includegraphics[width=0.475\textwidth]{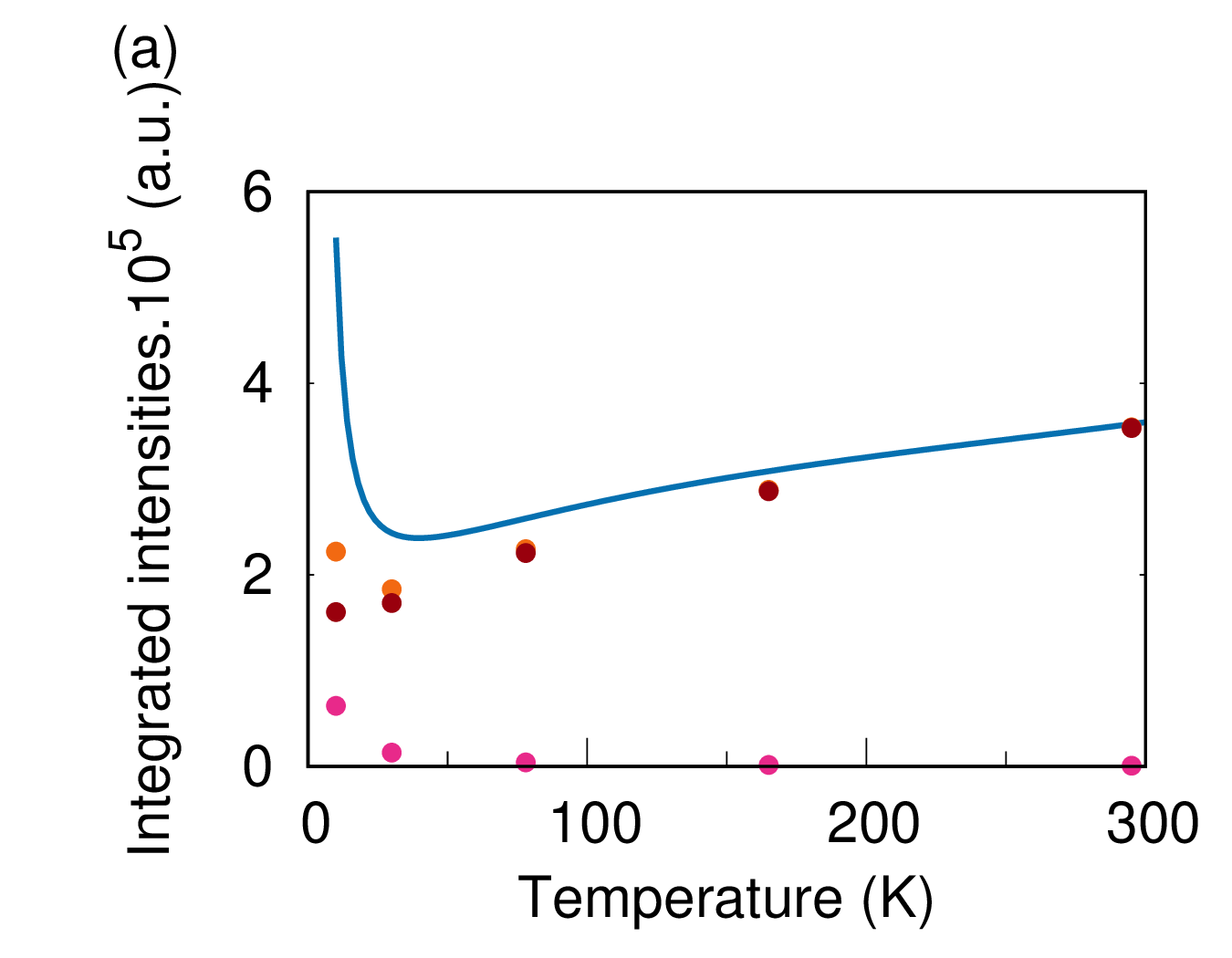}
         \includegraphics[width=0.475\textwidth]{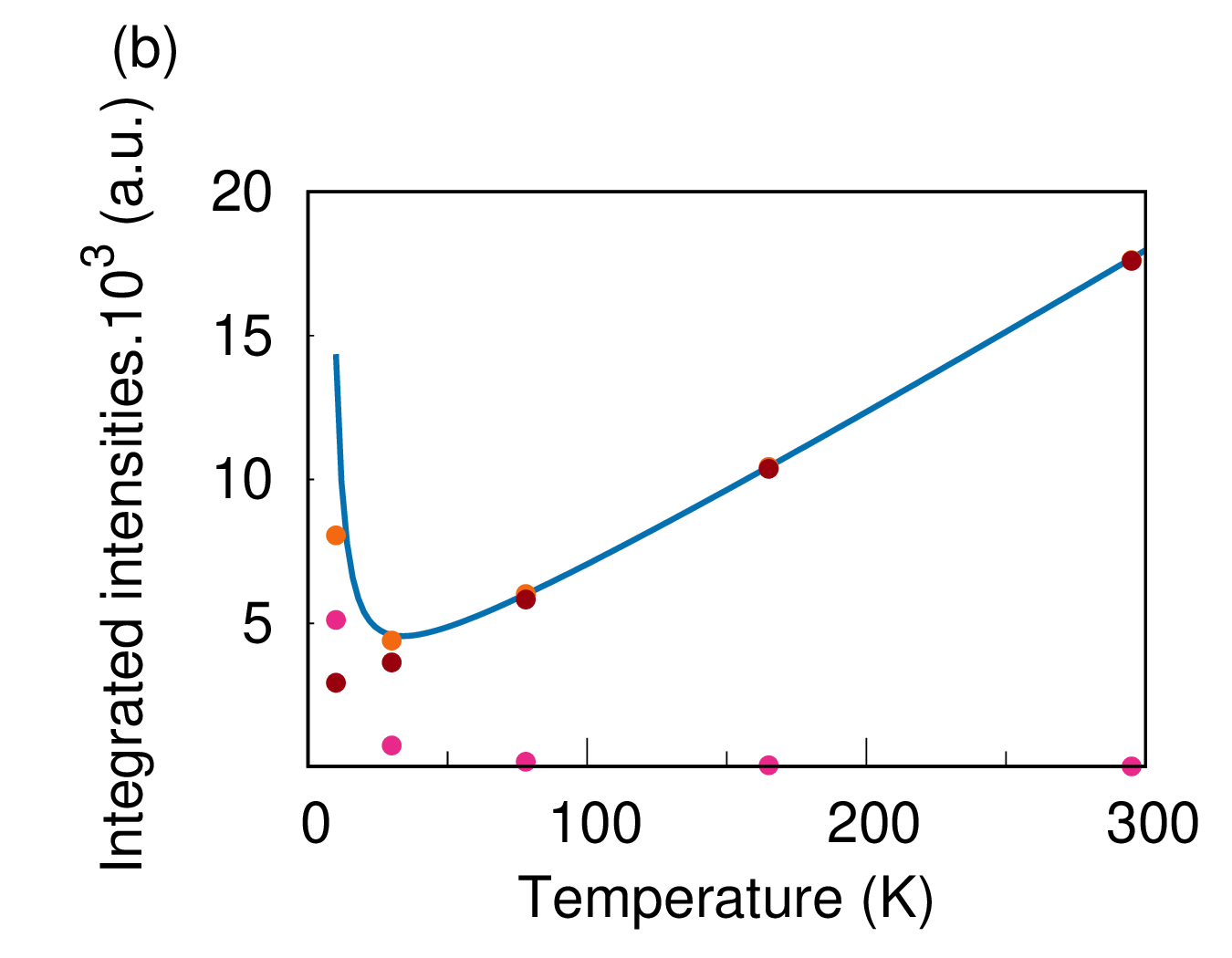}
         \includegraphics[width=0.475\textwidth]{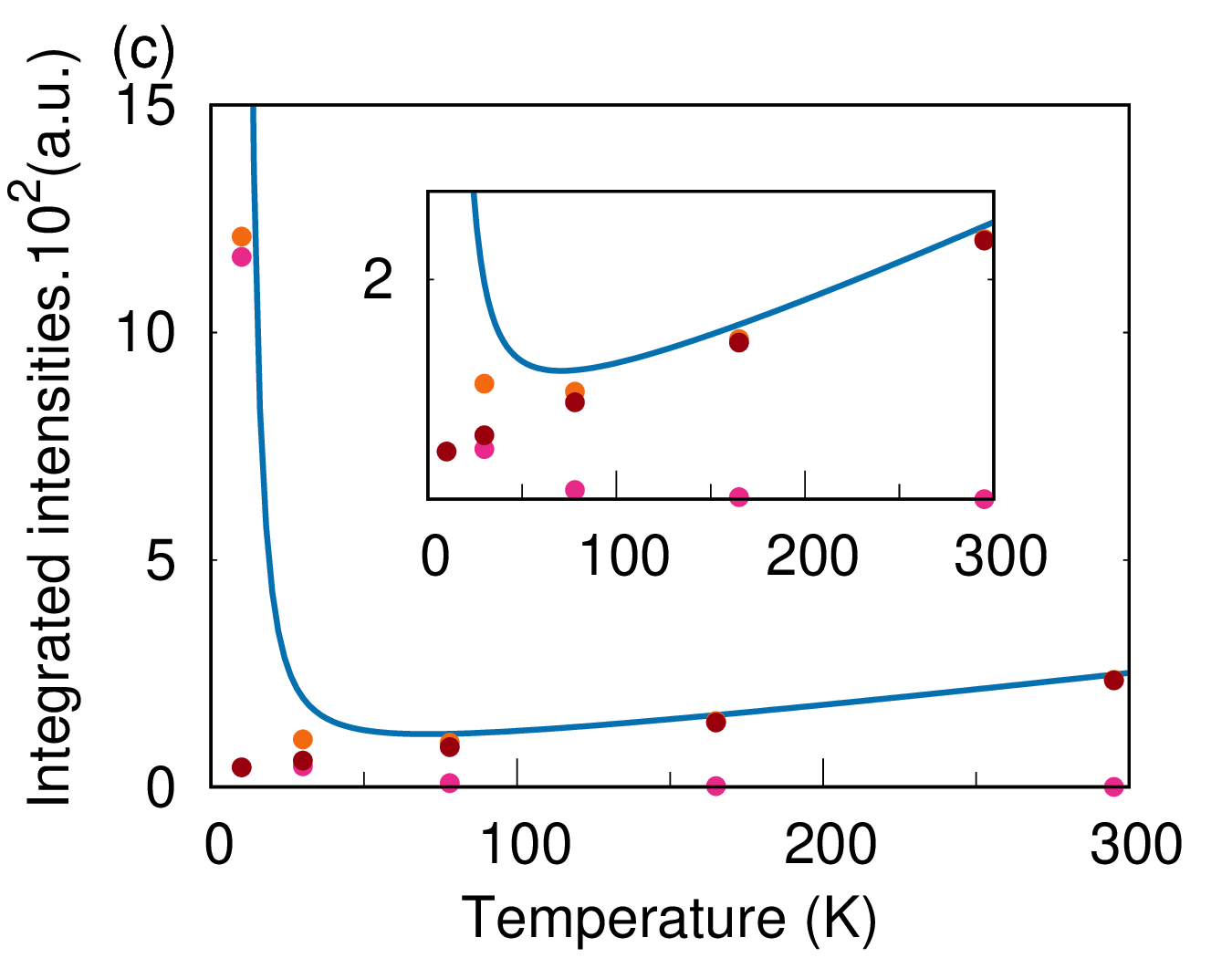}
         \includegraphics[width=0.475\textwidth]{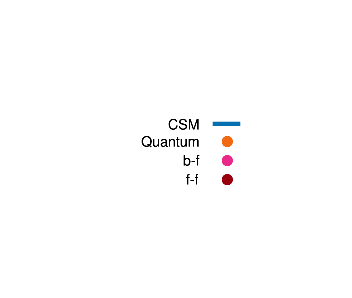}
         %\caption{CIA of NeAr}
         \caption{Integrated intensities at different temperatures calculated using equation(\ref{csm}) are shown as the solid line. Dots mark results of quantum mechanical lineshape calculations split out in contributions of free-to-free and bound-to-free transitions. Different panels correspond to (a) Ne--He (b) Ar--He and (c) Ar--Ne.}
         \label{int_ints}
\end{figure*}

The total spectral intensity arises from a combination of bound-bound, bound-free, and free-free transitions.
Quantifying the relative contributions of each transition type is critical to identifying the dominant transition mechanisms at a given temperature.
Using the classical statistical approach, we calculated the integrated intensity, which is then compared with the integrated intensities obtained from the spectra of free-free and bound-free transitions.
This comparison is shown in Fig.~\ref{int_ints}.
For the temperatures relevant for this study, the contribution from free-free transitions closely matches with the total intensity predicted by the classical model, indicating that free-free transitions are important as the continuum states become populated with increasing temperature \cite{vigasin_2004}. Contributions from bound-free transitions become significant only at very low temperatures (10 K). Among the systems studied, Ne--He and Ar--He have only one bound state each, while Ar--Ne has four bound states.  However, the contributions from bound-bound transitions are not included in the line-shape calculations, as they primarily become relevant only at very low temperatures and wavenumbers.

\begin{figure*} 
         \centering
         \includegraphics[width=0.475\textwidth]{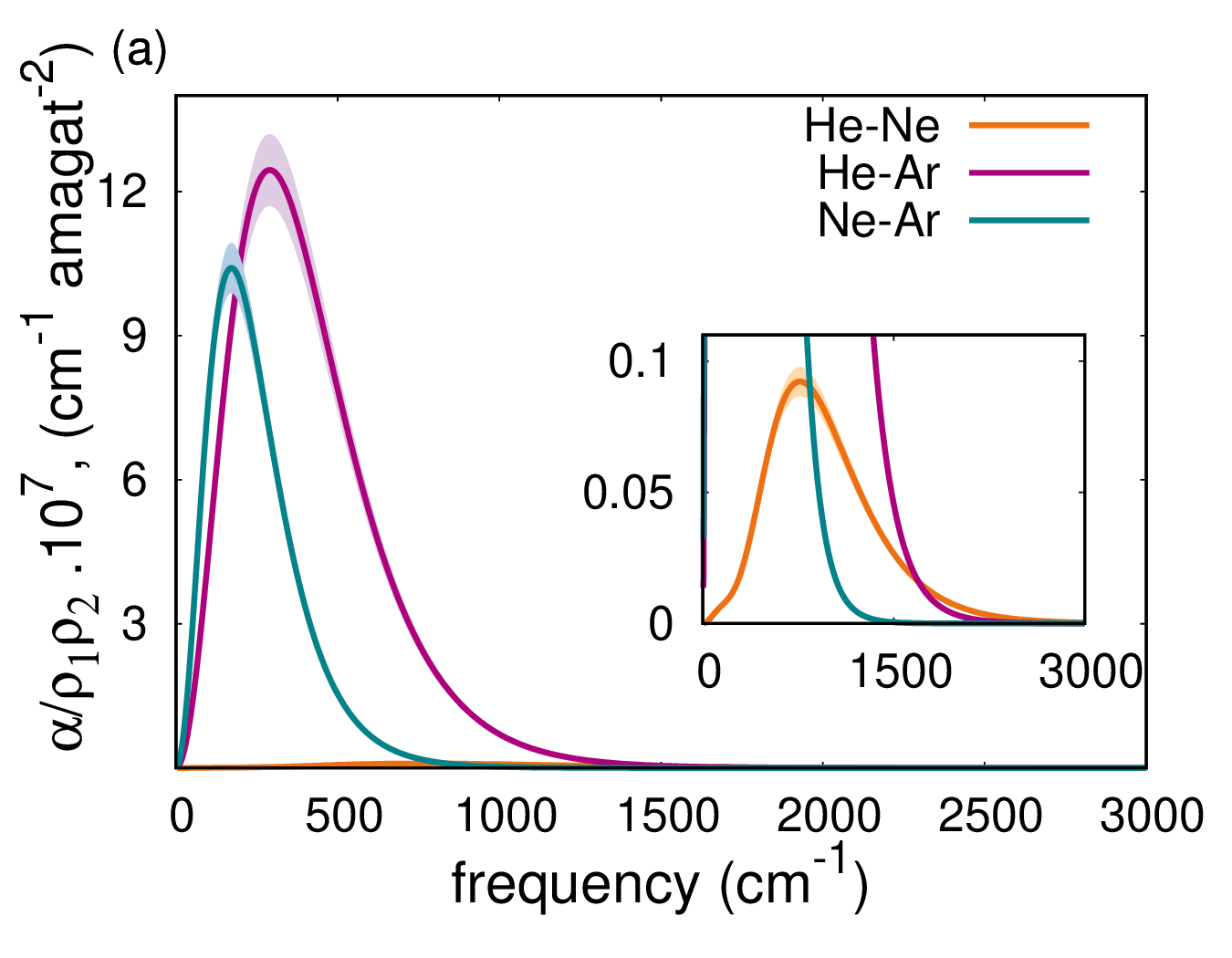}
         \includegraphics[width=0.475\textwidth]{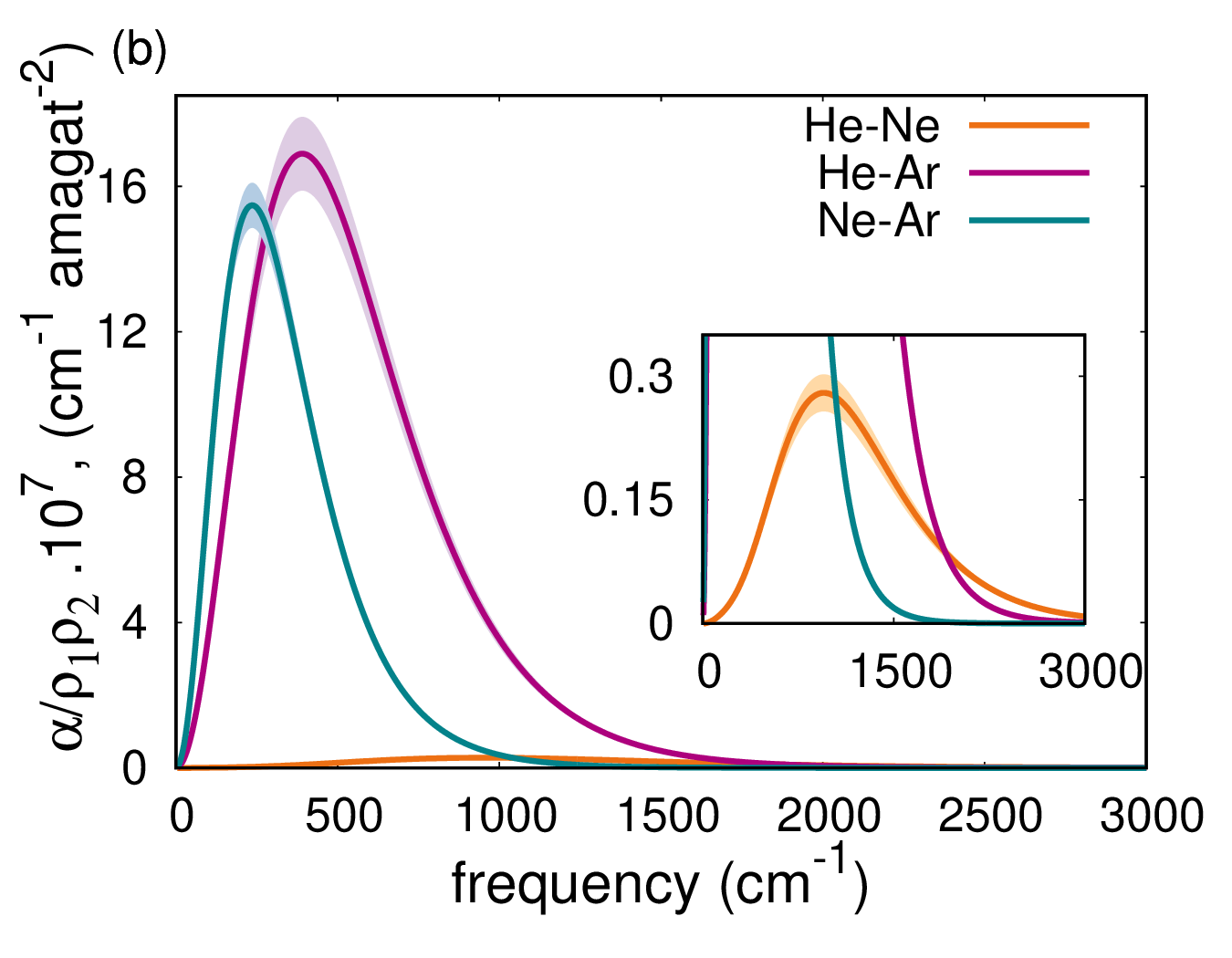}
         \caption{Predicted CIA spectra for noble gas atoms at high temperatures
                (a) 1000 K (b) 2000 K.}
         \label{CIA_high_temp}
\end{figure*}

CIA spectra for noble gas heteroatoms at high temperatures are shown in Figure \ref{CIA_high_temp}. The intensity of Ne--He is several orders of magnitude lower than Ar--He and Ar--Ne. This difference arises because of the small size of Ne--He, making them less polarizable, resulting in weaker dipole strength than Ar--He and Ar--Ne.
 
The observed widths of the spectral functions reflect the short lifetime of the fly-by interaction. The interaction time is proportional to the inverse of the speed of collision, and thus to the inverse of the reduced mass of the system. Therefore, a narrowing of the spectrum with increasing mass is observed. As the temperature increases, higher energy states become more populated, increasing the intensity of the spectra for a given system. Also, at high temperatures, the kinetic energies of the particles increase, resulting in broader spectra.

\section{CONCLUSIONS}

In this work, we have quantified for the first time to what extent calculated CIA spectra of noble gas atom pairs are sensitive to the quality of the underlying \emph{ab initio} data, particularly the potential energy and the induced dipole surfaces. 
We find that the spectra are more sensitive to the dipole surface than to the potential, and that the sensitivity decreases with temperature.
Even at cryogenic temperatures of 77~K where the sensitivity is greatest,
an overly conservative estimate of the error in the predicted spectra is in the worst case considered 43\% 
indicating that \emph{ab initio} calculations of CIA spectra can be predictive.
In many cases, and especially at higher temperature, the uncertainty can be substantially lower, which disagrees with frequently made claims that CIA spectra are highly sensitive probes of the potential energy and induced dipole surfaces \cite{elkader,el2014spectral,el2012rototranslational,BORYSOW1984235,meyer2015,fakhardji2019collision,borysow1998computer}.
Our analysis also shows that, for the rare gas mixtures, the absorption spectrum is never sensitive to interatomic distances where the induced dipole moment is determined by the van der Waals interaction, despite the extensive attention this has received in the literature\cite{bohr1,bohr2,fowler,heijmen1996symmetry,meyer1994long,bishop1993calculation,fowler1990d9,galatrygharbi,whisbye1,whisbye2,matchanesbet,lr2}.
Furthermore, our analysis reveals that the internuclear distance range contributing to spectra changes with temperature, highlighting the increasing role of short-range interactions at elevated temperatures. These findings refine our understanding of the temperature-dependent behavior of the spectra and provide clear guidance for improving the theoretical modeling for astronomical modeling.

For Ar--He and Ar--Ne, our results agree with experimental results to within our conservatively estimated error bars.
For these systems, our calculations provide absorption spectra for atmospheric modeling in an expanded temperature range.
For Ne--He mixtures, the agreement with scarce experimental data is less than satisfactory,
with previous experimental results well outside the estimated uncertainty of our calculations.
Furthermore, for this system, we predict an unknown and unexpected double-peak structure in the absorption spectrum.
A simple analytical model explains the lineshape results from the non-monotonic decay of the induced dipole function,
which cannot be described by commonly used simple single-exponential decay models of the dipole function.

The insensitivity of CIA spectra to the uncertainty in \emph{ab initio} potential and dipole surface, found here,
indicates that quantitative theoretical modeling is possible with accuracy better than needed for many astronomical applications.

\section{Acknowledgements}
EU Doctoral Network PHYMOL 101073474 (project call reference HORIZON-MSCA-2021-DN-01)

\begin{table}[h!]
    \centering
    \begin{tabular}{cccc}
     \hline
     \hline
      & Ne--He & Ar--He & Ar--Ne \\
     \hline
     diff PES & 4.22\%   & 3.54\%  & 4.46\% \\
     diff IDS&  1.41\% & 9.05\%  &  8.35\% \\
    \hline
    \hline
    \end{tabular}
    \caption{Relative difference between integrated intensity of spectra calculated using different PES and IDS computed in small AVTZ basis sets compared to the CBS limit at 295 K}
    \label{tab:diff_295K}
\end{table}

\begin{table}[h!]
    \centering
    \begin{tabular}{cccc}
     \hline
     \hline
      & Ne--He & Ar--He & Ar--Ne \\
     \hline
     diff PES & 3.06\%   & 2.07\%  & 3.32\% \\
     diff IDS&  2.77\% & 6.26\%  &  5.32\% \\
    \hline
    \hline
    \end{tabular}
    \caption{Relative difference between integrated intensity of spectra calculated using different PES and IDS computed in small AVTZ basis sets compared to the CBS limit at 2000 K}
    \label{tab:diff_2000K}
\end{table}

\begin{table}[h!]
    \centering
    \begin{tabular}{cccc}
     \hline
     \hline
     basis stes & Ne--He & Ar--He & Ar--Ne \\
     \hline
     AVTZ& 2.49\%   & 5.22\%  & 3.54\% \\
     AVQZ&  1.68\% & 3.35\%  &  1.79\% \\
     AV5Z&  1.56\% & 2.02\%  &  1.68\% \\
     AV6Z&  1.02\% & 1.19\%  &  1.08\% \\
    \hline
    \hline
    \end{tabular}
    \caption{Relative difference in CIA calculated using different basis sets compared to the CBS limit at 295 K}
    \label{tab:basis_295}
\end{table}

\begin{table}[h!]
    \centering
    \begin{tabular}{cccc}
     \hline
     \hline
     basis stes & Ne--He & Ar--He & Ar--Ne \\
     \hline
     AVTZ& 5.78\%   & 4.09\%  & 1.86\% \\
     AVQZ&  3.24\% & 2.92\%  &  1.21\% \\
     AV5Z&  1.63\% & 1.94\%  &  1.17\% \\
     AV6Z&  0.98\% & 1.16\%  &  0.75\% \\
    \hline
    \hline
    \end{tabular}
    \caption{Relative difference in CIA calculated using different basis sets compared to the CBS limit at 2000 K.}
    \label{tab:basis_2000}
\end{table}

\begin{table}[h!]
    \centering
    \begin{tabular}{cccc}
     \hline
     \hline
      & Ne--He & Ar--He & Ar--Ne \\
     \hline
     CCSD & 9.74\%   & 9.66\%  & 12.84\% \\
     CCSD(T)&  0.61\% & 0.51\%  &  0.46\% \\
    \hline
    \hline
    \end{tabular}
    \caption{Relative difference between integrated intensity of spectra calculated using the CCSD and CCSD(T) methods compared to the estimated FCI limit at 295K.}
    \label{tab:method_295}
\end{table}

\begin{table}[h!]
    \centering
    \begin{tabular}{cccc}
     \hline
     \hline
      & Ne--He & Ar--He & Ar--Ne \\
     \hline
     CCSD & 0.98\%   & 5.159\%  & 4.67\% \\
     CCSD(T)&  0.24\% & 0.28\%  &  0.19\% \\
    \hline
    \hline
    \end{tabular}
    \caption{Relative difference between integrated intensity of spectra calculated using the CCSD and CCSD(T) methods compared to the estimated FCI limit at 2000 K.}
    \label{tab:method_2000}
\end{table}

\begin{figure*} 
         \centering
         \includegraphics[width=0.475\textwidth]{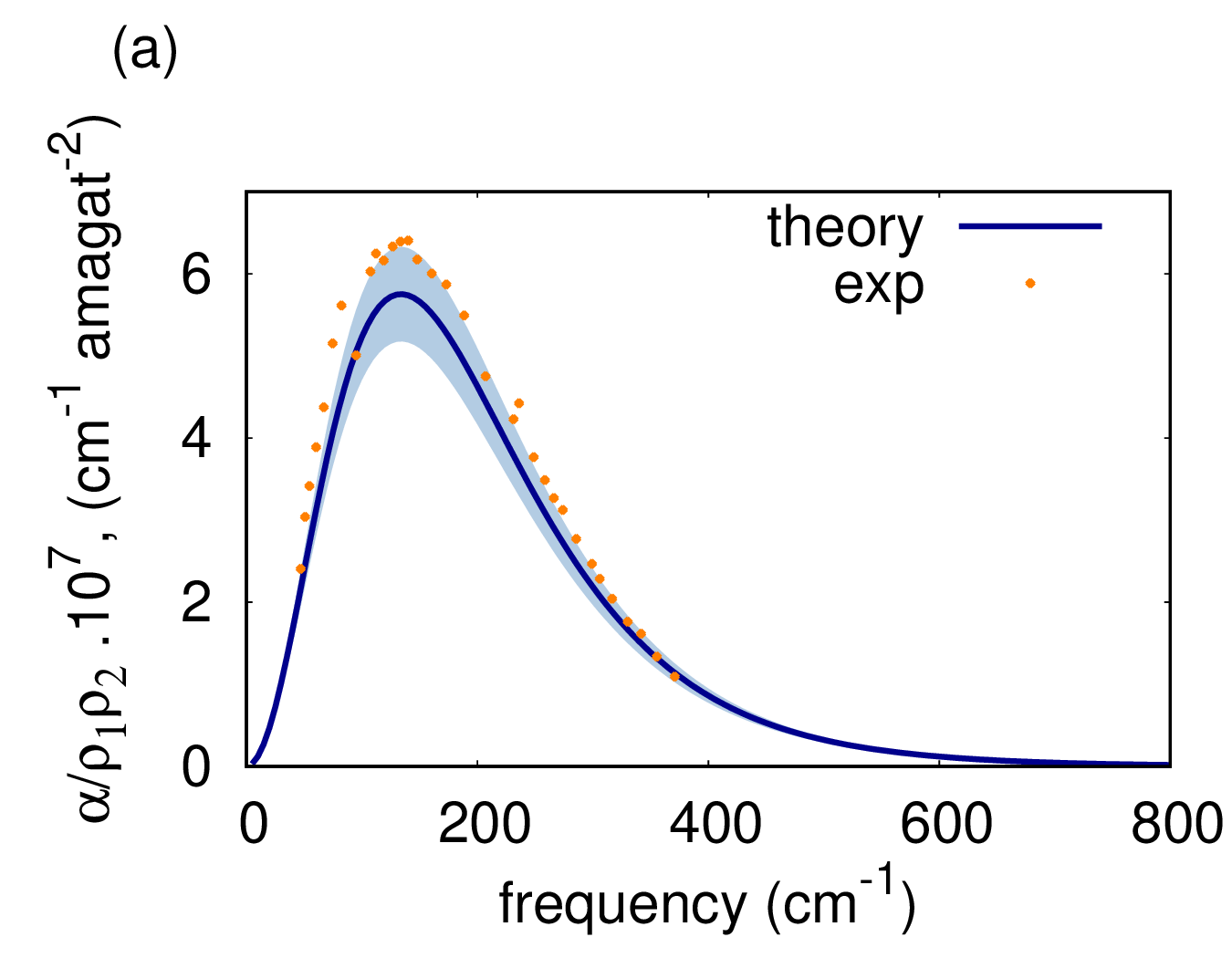}
         \includegraphics[width=0.475\textwidth]{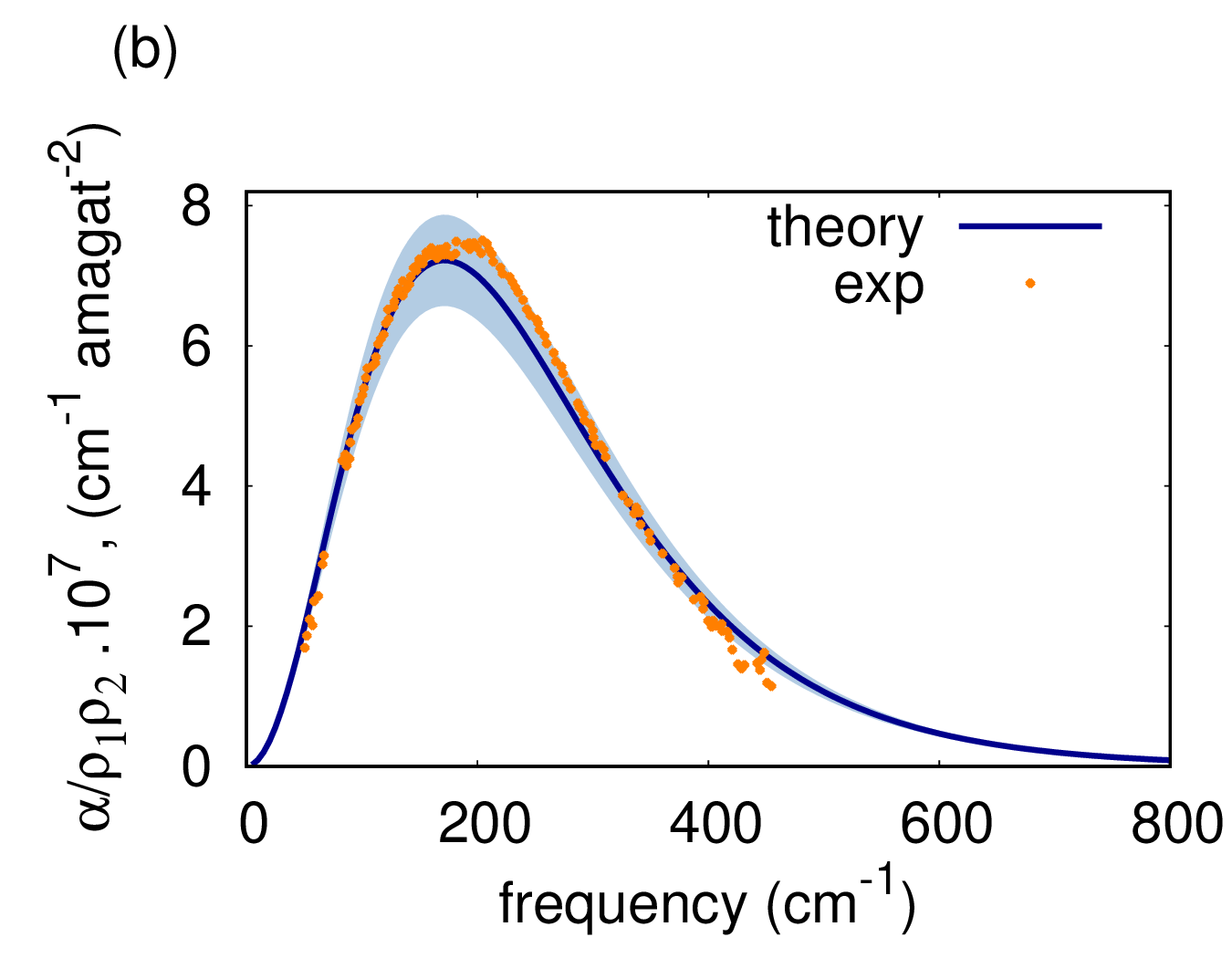}
         \includegraphics[width=0.475\textwidth]{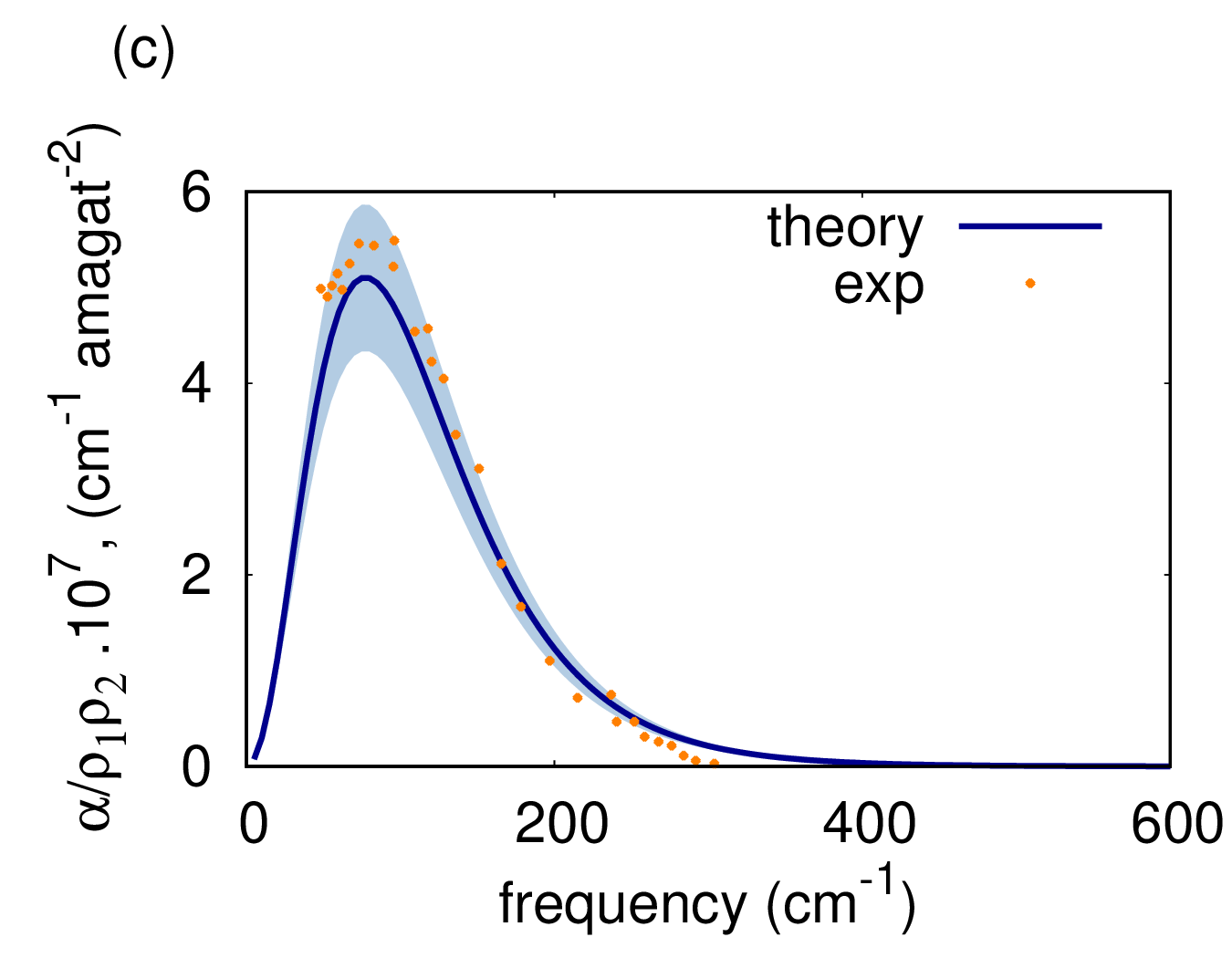}
         \includegraphics[width=0.475\textwidth]{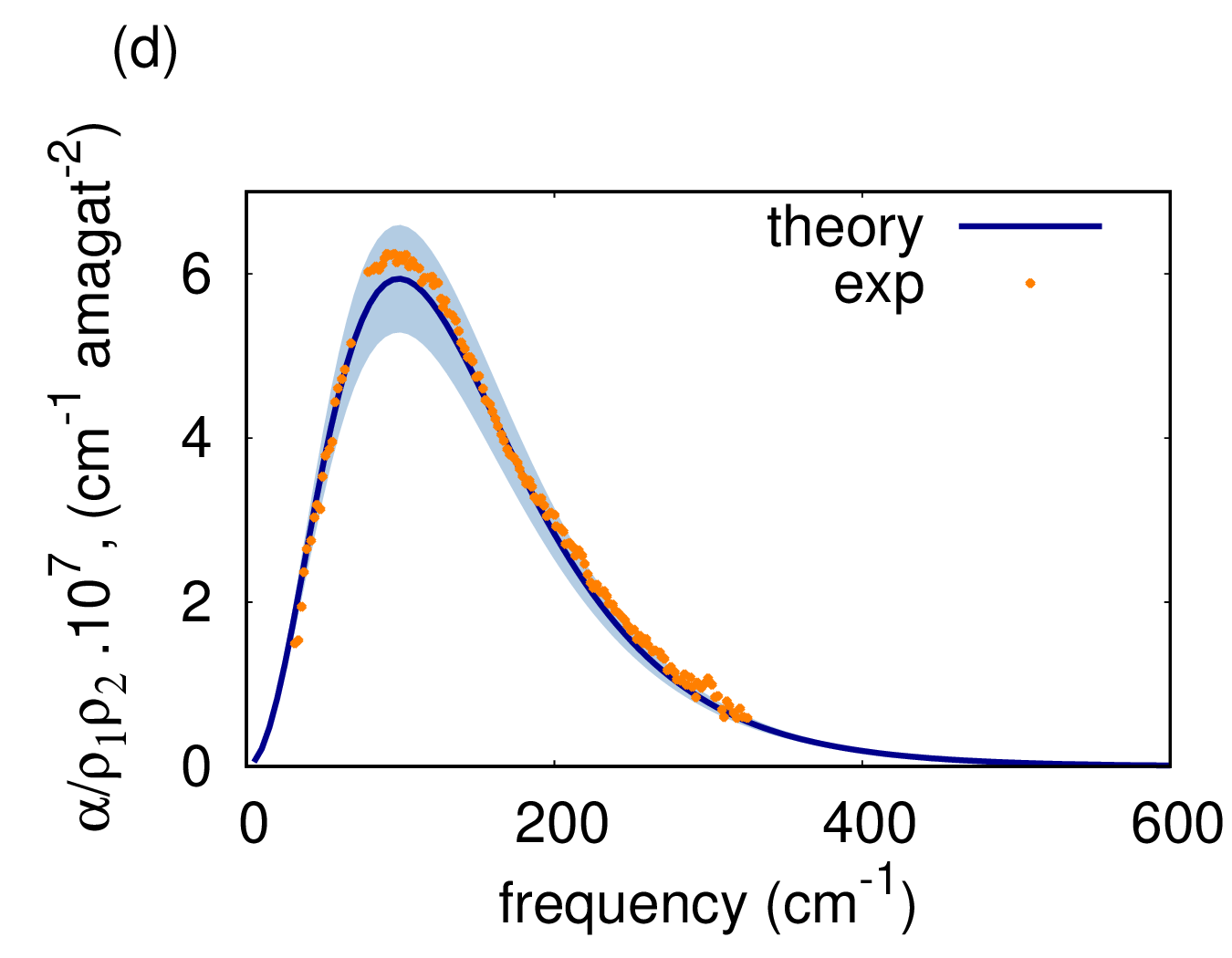}
         \includegraphics[width=0.475\textwidth]{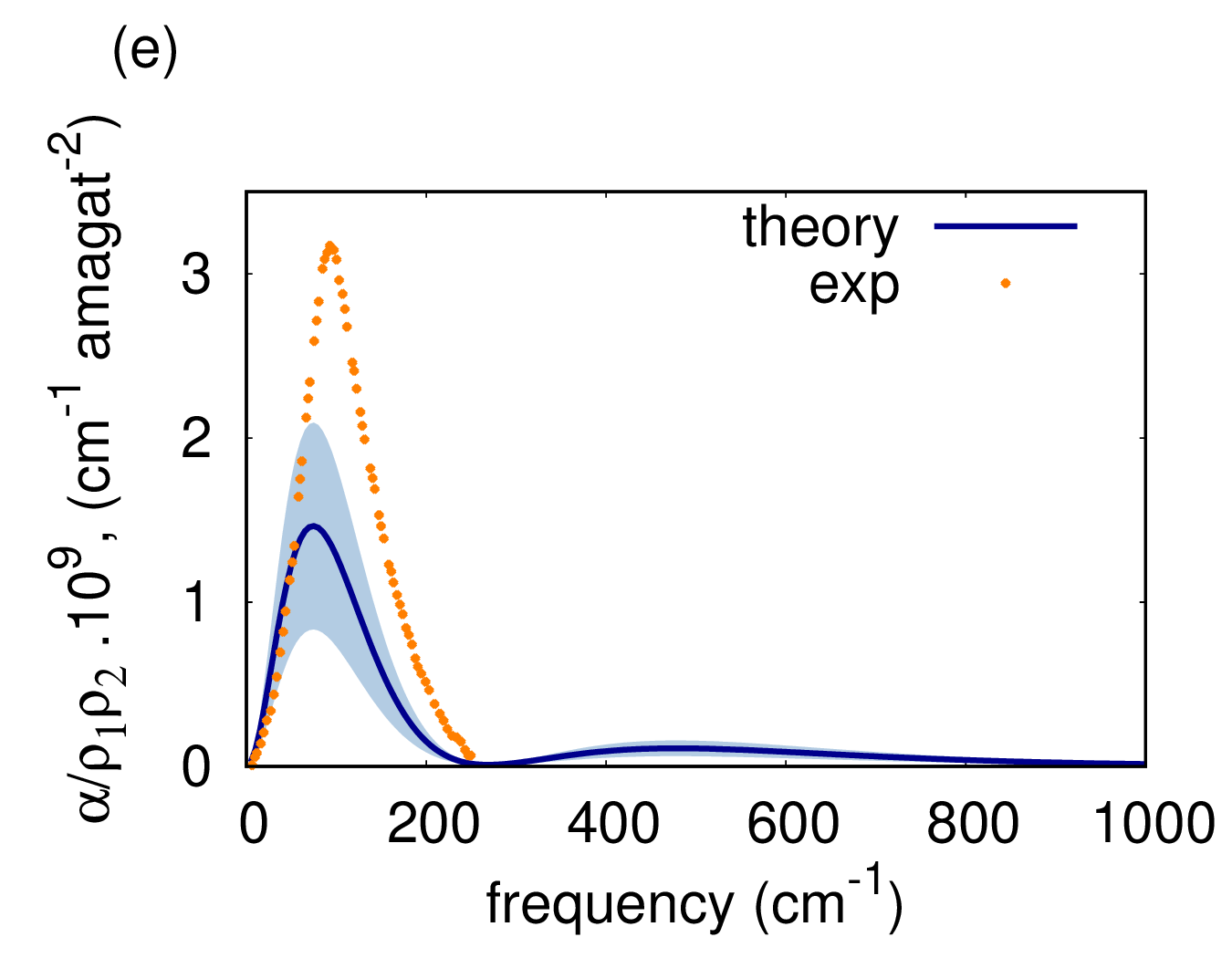}
         \caption{Comparison between experimental measurements and theoretical calculations. (a) CIA of Ar--He at 165 K (b) CIA of Ar--He at 295 K 
         (c) CIA of Ar--Ne at 165 K (d) CIA of Ar--Ne at 295 K (e) CIA of Ne--He at 77 K. Blue line is the calculated spectra, orange dots are the experimental values from \cite{CIA_exp,hene_exp,CIA_exp_bw1,bt1,bt2,bt3} and the shaded region represents the error bar. }
         \label{theory_vs_exp}
\end{figure*}

\newpage

\end{document}